\documentclass[epj]{svjour}
\usepackage{amsmath}
\usepackage{graphicx}
\usepackage{hyperref}
\usepackage{subfigure}
\usepackage{color}

\begin{document}

\title{Triangle singularity in the $J/\psi \rightarrow K^+ K^- f_0(980)(a_0(980))$ decays}

\author{Wei-Hong Liang\inst{1}
\and Hua-Xing Chen\inst{2}
\thanks{hxchen@buaa.edu.cn}
\and Eulogio Oset\inst{1,3}
\thanks{eulogio.oset@ific.uv.es}
\and En Wang\inst{4}
\thanks{wangen@zzu.edu.cn}
}                     
\offprints{}          
\institute{
Department of Physics, Guangxi Normal University, Guilin 541004, China
\and
School of Physics, Beihang University, Beijing 100191, China
\and
Departamento de F\'{\i}sica Te\'orica and IFIC, Centro Mixto Universidad de Valencia-CSIC Institutos de Investigaci\'on de Paterna, Aptdo. 22085, 46071 Valencia, Spain
\and
School of Physics and Engineering, Zhengzhou University, Zhengzhou, Henan 450001, China
}
\date{Received: date / Revised version: date}
%
\abstract{
We study the $J/\psi \rightarrow K^+ K^- f_0(980)(a_0(980))$ reaction and find that the mechanism to produce this decay develops a triangle singularity around $M_{\rm inv}(K^- f_0/K^- a_0) \approx 1515$~MeV. The differential width $d\Gamma / dM_{\rm inv}(K^- f_0/K^- a_0)$ shows a rapid growth around the invariant mass being 1515~MeV as a consequence of the triangle singularity of this mechanism, which is directly tied to the nature of the $f_0(980)$ and $a_0(980)$ as dynamically generated resonances from the interaction of pseudoscalar mesons. The branching ratios obtained for the $J/\psi \rightarrow K^+ K^- f_0(980)(a_0(980))$ decays are of the order of $10^{-5}$, accessible in present facilities, and we argue that their observation should provide relevant information concerning the nature of the low-lying scalar mesons.
} 
\maketitle

\section{Introduction}
\label{sec:intro}

\begin{figure*}[hbt]
\begin{center}
\subfigure[]{
\includegraphics[width=0.4\textwidth]{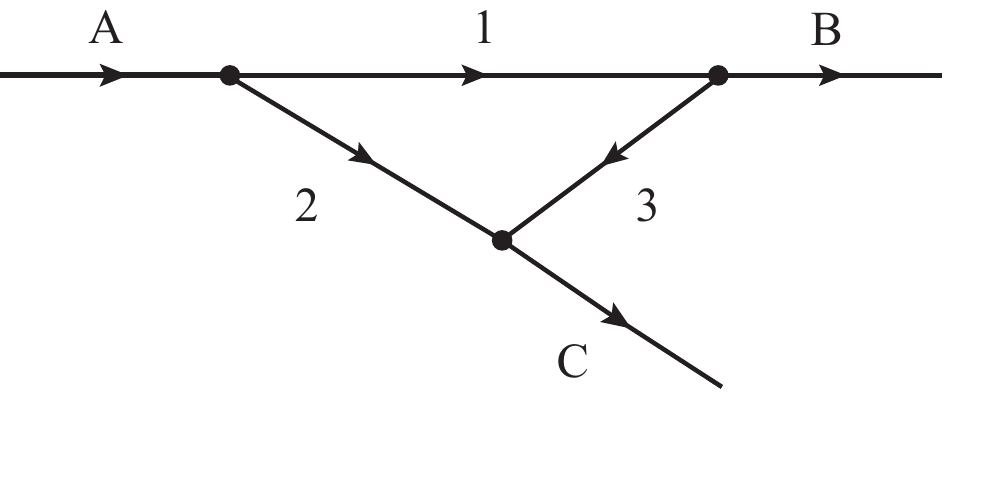}}
~~~~~~~~~~
\subfigure[]{
\includegraphics[width=0.4\textwidth]{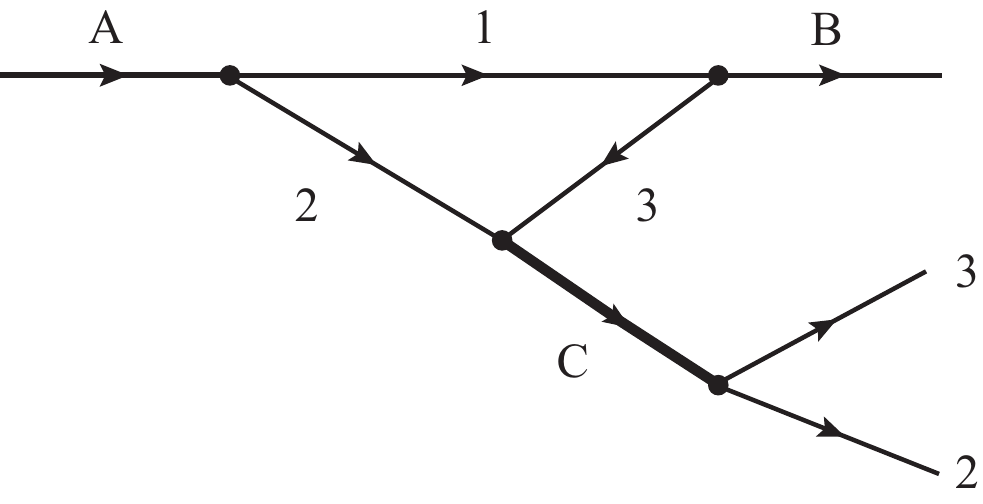}}
\caption{Mechanism for a triangle singularity.}
\label{fig:example}
\end{center}
\end{figure*}

Discussed already in Ref.~\cite{Karplus:1958zz}  and systematized by Landau in Ref.~\cite{Landau:1959fi}, the triangle singularities (TS) were fashionable in the sixties~\cite{Peierls:1961zz,Aitchison:1964zz,Chang,Bronzan:1964zz,Coleman:1965xm,Schmid:1967ojm} and efforts were done to understand some reactions through TS mechanisms~\cite{Booth:1961zz,Anisovich}. A triangle mechanism stems from the decay of a particle $A$ into $1+2$, followed by the decay of $1$ into $3+B$, and posterior fusion of $2+3$ to give a new particle $C$ (see Fig.~\ref{fig:example}(a)) or $2+3$ (see Fig.~\ref{fig:example}(b), rescattering), or a different pair of particles. It was shown in Ref.~\cite{Landau:1959fi} that when all these particles, $1,2,3$, can be placed on shell in the corresponding Feynman diagram, a singularity can develop in the corresponding amplitude. The conditions for the singularity are made more specific by Coleman-Norton~\cite{Coleman:1965xm} showing that particles $1$ and $B$ have to be parallel in the $A$ rest frame and the process has to be possible at the classical level. Analytical expressions of these conditions can be see in Ref.~\cite{Liu:2015taa} and in a simpler form in Ref.~\cite{Bayar:2016ftu}. The formalism of Ref.~\cite{Bayar:2016ftu} allows one to see the explicit effect of the width of particle $1$ in the shape of the singularity, and this is explicitly shown in Ref.~\cite{Debastiani:2018xoi} where some considerations are made concerning the Schmid theorem~\cite{Schmid:1967ojm}, which states that in the case of the mechanism with $2 + 3 \rightarrow 2 + 3$ (rescattering) the triangle singularity can be absorbed by the tree level diagram $A \rightarrow 1 + 2~(1 \rightarrow 3 + B)$. In Ref.~\cite{Debastiani:2018xoi} it is shown that this only occurs in the limit of zero width for particle $1$, where the triangle mechanism is negligible with respect to the tree level one.

%
%

With the advent of a large amount of experimental information, examples of triangle singularities have become available lately, and the field has experienced a revival. The spark was raised by the puzzle of the anomalously large isospin breaking in the $\eta(1405) \rightarrow \pi^0 f_0(980)$ reaction~\cite{BESIII:2012aa}, which was explained in Refs.~\cite{Wu:2011yx,Aceti:2012dj} in terms of a triangle singularity (see also following references~\cite{Achasov:2015uua,Achasov:2018swa}). It is interesting to recall the mechanism of Refs.~\cite{Wu:2011yx,Aceti:2012dj}, which has served to disentangle related reactions and to make predictions for new reactions that should see TS effects.
The mechanism studied in Refs.~\cite{Wu:2011yx,Aceti:2012dj} consists of $\eta(1405) \rightarrow K^* \bar K$, followed by $K^* \rightarrow \pi K$ and later fusion of $K \bar K$ to give $a_0(980)$ or $f_0(980)$. There is no problem in having $K \bar K \rightarrow a_0(980)$, since $\pi a_0(980)$ can match to zero isospin of the $\eta(1405)$, but $K \bar K \rightarrow f_0(980)$ is isospin forbidden, since $\pi f_0(980)$ has isospin one. The difference of masses between $K^+ K^-$ and $K^0 \bar K^0$ in the triangle loop prevents the exact cancellation of two diagrams that occurs in the limit of equal masses where isospin is conserved. This also leads to a peculiar very narrow shape of the $f_0(980)$ that is not tied to the natural width of the $f_0(980)$ but to the difference of masses between $K^+ K^-$ and $K^0 \bar K^0$.

The above reaction is also very enlightening from the point of view of the nature of the $f_0(980)$ and $a_0(980)$, since these resonances are not directly produced but come from the rescattering of the $K \bar K$ components, {\it i.e.}, the mechanism for the formation of these resonances in the chiral unitary approach~\cite{Oller:1997ti,Kaiser:1998fi,Locher:1997gr,Nieves:1999bx}. This same TS mechanism appears in the $\tau^- \to \nu_\tau \pi^- f_0(a_0)$ decay~\cite{Dai:2018rra}, 
the $B_s^0  \rightarrow J/\psi \pi^0  f_0(a_0)$ decay~\cite{Liang:2017ijf}, the $D_s^+ \to \pi^+ \pi^0 f_0(a_0)$ decay~\cite{Sakai:2017iqs}, and the $B^-\rightarrow D^{*0} \pi^- f_0(a_0)$ decay~\cite{Pavao:2017kcr}.
The same triangle singularity was shown in Refs.~\cite{Ketzer:2015tqa,Aceti:2016yeb} to provide a natural explanation for the peak around 1420 MeV observed in the $\pi f_0(980)$ final state in diffractive $\pi p$ collisions by the COMPASS Collaboration, which was originally branded as a new ``$a_1(1420)$'' resonance~\cite{Adolph:2015pws}. It is easy to envisage many reactions of this type, one also from the $J/\psi$ decay, as $J/\psi \to \pi_{\tiny\mbox{\textcircled{1}}} K^* \bar K$ followed by $K^* \to K \pi_{\tiny\mbox{\textcircled{2}}}$ and $K \bar K \to f_0(980)$. Once again one can anticipate a peak at $M_{\rm inv}( \pi_{\tiny\mbox{\textcircled{2}}} f_0) \approx 1420$~MeV like in the other reactions. However, the possibility to have different pairs from the $\pi^+ \pi^- \pi^+ \pi^-$ final state forming the $f_0(980)$ makes the experimental analysis and the theoretical work more involved.

Related reactions, $\tau^- \to \nu_\tau \pi A$ ($A$ for axial-vector mesons)~\cite{Dai:2018zki} and concretely the $\tau^- \to \nu_\tau \pi^- f_1(1285)$~\cite{Oset:2018zgc}, rely upon the $K^* \bar K^* K$ intermediate states with a good description of the $\tau^- \to \nu_\tau \pi^- f_1(1285)$ data~\cite{pdg} (see alternative approach in Ref.~\cite{Volkov:2018fyv} based on the Nambu-Jona-Lasinio model). For these latter reactions, the $\bar K^* K$ in the triangle loop fuse to give the axial vector mesons, which are also dynamically generated from the pseudoscalar-vector mesons interactions according to the chiral unitary approach~\cite{Lutz:2003fm,Roca:2005nm,Zhou:2014ila}.

A related mechanism is also used in Ref.~\cite{Liu:2017vsf} to describe the $B_c\to B_s \pi\pi$ reaction with the $1,2,3$ particles being $\bar K^* B \bar K$, the $B^- \rightarrow K^- \pi^- D_{s0}^{*+}(2317)$ reaction~\cite{Sakai:2017hpg} with $K^* D K$ in the intermediate states, the $\Lambda_c\rightarrow\pi\pi\Lambda(1405)$ reaction~\cite{Dai:2018hqb} with $\bar K^* p \bar K$ in the intermediate states, and the $\Lambda_c \to \pi \pi \Sigma^*$ reaction~\cite{Xie:2018gbi} with $\bar K^* p \bar K$ in the intermediate states. Other reactions rely on very different intermediate states, like in the $\Psi \to K \bar K J/\psi$ reaction~\cite{Cao:2017lui} with $D_{s1} \bar D_s D^*$ in the intermediate states, or the $\Lambda_c \to \pi \phi p$ reaction~\cite{Xie:2017mbe} with $\Sigma^* K^* \Sigma$ in the intermediate states. A recent review of reactions explained in terms of TS mechanisms and predictions made for many other physical processes can be seen in Ref.~\cite{Debastiani:2018xoi}.

\begin{figure*}[hbt]
\begin{center}
\subfigure[]{
\includegraphics[width=0.4\textwidth]{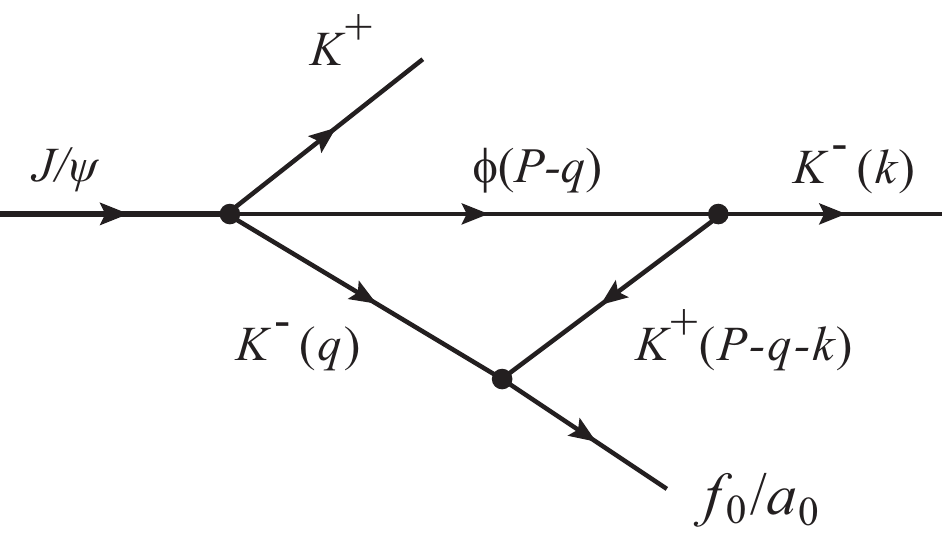}}
~~~~~~~~~~
\subfigure[]{
\includegraphics[width=0.4\textwidth]{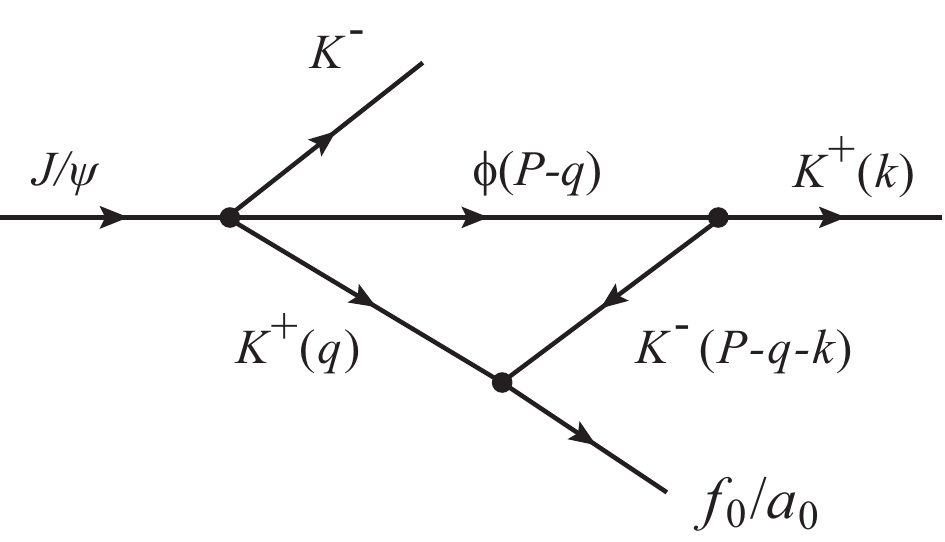}}
\caption{Triangle diagrams for the $J/\psi \rightarrow K^+ K^- f_0(980)(a_0(980))$ decays. The parentheses give the four-momenta of the particles with (a) $P = p_{J/\psi} - p_{K^+}$ and (b) $P = p_{J/\psi} - p_{K^-}$.}
\label{fig:triangle}
\end{center}
\end{figure*}

The reaction proposed here relies upon a different intermediate state, not discussed previously, which involves the $K \bar K \phi$ intermediate states, as depicted in Fig.~\ref{fig:triangle}. The reaction is $J/\psi \rightarrow K^+ K^- f_0(980)(a_0(980))$. It involves the strong $J/\psi$ decay and the mechanism proceeds via $J/\psi \rightarrow K^+ K^- \phi$, followed by $\phi \to K^- K^+$ and the posterior fusion of $K^- K^+$ to give either the $f_0(980)$ or $a_0(980)$. Both reactions are possible in the present case but the rates of production are tied to the way the $K^+ K^-$ generates the $f_0(980)$ or $a_0(980)$ resonance through its interaction, hence, providing new information on the nature of these resonances. The intermediate states in the loop, $1,2,3$, are now $\phi K^- K^+$. In addition, the $\phi$ is very narrow, $\Gamma_\phi = 4.25$~MeV~\cite{pdg}, and in particular the TS structure should be narrow around the TS point given by the equation~\cite{Bayar:2016ftu}
\begin{equation}
q_{\rm on} - q_{a-} = 0 \, ,
\label{con:triangle}
\end{equation}
where $q_{\rm on}$ is the on-shell momentum of the $\phi$ and $q_{a-}$ the on-shell $K^-$ momentum in the loop, antiparallel to the $\phi$.

%
%

It should be noted that the TS condition of Eq.~(\ref{con:triangle}) is now fulfilled only in a very narrow window of $K^+ K^-$ energies between 987 MeV and 993 MeV, where the $f_0(980)$ and $a_0(980)$ peak. Away but close to the point where the TS appears, the amplitudes are no longer singular in the $\Gamma_\phi \rightarrow 0$ limit, but a peak structure still remains by inertia. Yet, this feature confers the amplitude a special signature that makes the shapes different to ordinary cases of $f_0(980)$ or $a_0(980)$ production, which is tied again to the dynamical nature of these resonances formed from the pseudoscalar-pseudoscalar interaction in coupled channels. The reaction, hence, contains relevant information concerning the nature of the $f_0(980)$ and $a_0(980)$ resonances.

The condition of Eq.~(\ref{con:triangle}) is very useful to know when one has a TS, and helps rule out related triangle mechanisms which however do not develop a singularity, and hence do not compete with the singular mechanisms. In this sense we can envisage a primary $J/\psi \rightarrow \bar K K^* \pi$ decay with $K^* \to K \pi$ and the two pions merging into the $f_0(980)$. We can apply the condition of Eq.~(\ref{con:triangle}) to this mechanism containing $K^* \pi \pi$ in the loop, but with the same final $\bar K K f_0(980)$ state as in the mechanism of Fig.~\ref{fig:triangle}. However, in this case one can see that $q_{\rm on}$ and $q_{a-}$ are very far from each other, leading to a mechanism that cannot compete with the singular one, and which in any case only provides a smooth background in the region where the mechanism of Fig.~\ref{fig:triangle} produces a peak, which is what we want to investigate.
The smaller coupling of $f_0(980)$ to $\pi \pi$ than to $K \bar K$ also helps to make this background smaller.

\section{Formalism}
\label{sec:formalism}

Our mechanism for the $J/\psi \rightarrow K^+ K^- f_0(980)(a_0(980))$ reaction is depicted in Fig.~\ref{fig:triangle}. There is a primary decay of $J/\psi \rightarrow K^+ \phi K^-$, a second decay of $\phi \to K^- K^+$, and the posterior fusion of $K^- K^+$ to produce the $f_0(980)$ or $a_0(980)$ resonance. Given the complicated dynamics of the whole process, our strategy to provide absolute numbers for the decay width and mass distributions consists of taking the information for the primary step $J/\psi \rightarrow K^+ \phi K^-$ from experiment and the rest can be calculated reliably. In view of this, we study in a first step the reaction $J/\psi \rightarrow K^+ \phi K^-$.

\subsection{The $J/\psi \rightarrow K^+ K^- \phi$ decay}
\label{sec:tree}

The $J/\psi \rightarrow K^+ K^- \phi$ decay can proceed via $S$-wave. 
We take its amplitude, suited to the production of two vectors, as in Refs.~\cite{Pavao:2017kcr,Sakai:2017hpg}:
\begin{equation}
t_{J/\psi \rightarrow K^+ K^- \phi} = \mathcal{A}~{\vec \epsilon}(J/\psi) \cdot {\vec \epsilon}(\phi) \, ,
\label{eq:tree1}
\end{equation}
where ${\vec \epsilon}(J/\psi)$ and ${\vec \epsilon}(\phi)$ are the polarization vectors of the $J/\psi$ and $\phi$, respectively. Then we write the $K^- \phi$ invariant mass distribution of this decay as
\begin{eqnarray}
&& {d \Gamma_{J/\psi \rightarrow K^+ K^- \phi} \over d M_{\rm inv}(K^- \phi)}
\label{eq:tree2}
\\ \nonumber &=& {1 \over (2\pi)^3} {1 \over 4 M_{J/\psi}^2}~p_{K^+} \widetilde p_{K^-}~\overline{\sum} \sum \left|t_{J/\psi \rightarrow K^+ K^- \phi}\right|^2 \, ,
\end{eqnarray}
where $p_{K^+}$ is the momentum of the $K^+$ in the $J/\psi$ rest frame, and $\widetilde p_{K^-}$ is the momentum of the $K^-$ in the $K^- \phi$ rest frame:
\begin{eqnarray}
\nonumber p_{K^+} &=& {\lambda^{1/2}(M_{J/\psi}^2, m_{K^+}^2, M^2_{\rm inv}(K^- \phi)) \over 2 M_{J/\psi} }  \, ,
\\ \widetilde p_{K^-} &=& {\lambda^{1/2}(M^2_{\rm inv}(K^- \phi), m_{K^-}^2, m_{\phi}^2 ) \over 2 M_{\rm inv}(K^- \phi) } \, .
\end{eqnarray}
Here $\lambda(x, y, z)$ is the K{\"a}llen function $\lambda(x, y, z) = x^2 + y^2 + z^2 - 2xy - 2yz - 2xz$.

After summing over polarizations of Eq.~(\ref{eq:tree1}),
\begin{eqnarray}
&& \overline{\sum} \sum \left|t_{J/\psi \rightarrow K^+ K^- \phi}\right|^2
\nonumber \\ &=& \sum {1\over3}~\mathcal{A}^2~{\vec \epsilon}(J/\psi) \cdot {\vec \epsilon}(\phi)~{\vec \epsilon}(J/\psi) \cdot {\vec \epsilon}(\phi)
\nonumber \\ &=& \mathcal{A}^2 \, ,
\end{eqnarray}
we can simplify Eq.~(\ref{eq:tree2}) to be
\begin{equation}
{d \Gamma_{J/\psi \rightarrow K^+ K^- \phi} \over d M_{\rm inv}(K^- \phi)} = {1 \over (2\pi)^3} {1 \over 4 M_{J/\psi}^2}~p_{K^+} \widetilde p_{K^-}~\mathcal{A}^2 \, .
\end{equation}
Thus, the branching ratio of the $J/\psi \rightarrow K^+ K^- \phi$ decay can be calculated through
\begin{eqnarray}
\nonumber && {\rm Br}(J/\psi \rightarrow K^+ K^- \phi)
\\ \nonumber &=& {1 \over \Gamma_{J/\psi}} \int{{d \Gamma_{J/\psi \rightarrow K^+ K^- \phi} \over d M_{\rm inv}(K^- \phi)} d M_{\rm inv}(K^- \phi)}
\\ &=& {\mathcal{A}^2 \over \Gamma_{J/\psi}} \int {1 \over (2\pi)^3} {1 \over 4 M_{J/\psi}^2}~p_{K^+} \widetilde p_{K^-}~d M_{\rm inv}(K^- \phi) \, ,
\end{eqnarray}
where the integration is performed from $M_{\rm inv}(K^- \phi)_{\rm min} = m_{K^-} + m_\phi$ to $M_{\rm inv}(K^- \phi)_{\rm max} = m_{J/\psi} - m_{K^+}$.

The branching ratio of the $J/\psi \rightarrow K^+ K^- \phi$ decay has been experimentally measured to be~\cite{pdg}
\begin{eqnarray}
{\rm Br}(J/\psi \rightarrow K^+ K^- \phi) &=& (8.3\pm1.2)\times10^{-4} \, .
\label{experiment}
\end{eqnarray}
Using this as the input, we can determine the value of the constant $\mathcal{A}$, satisfying:
\begin{eqnarray}
\nonumber {\mathcal{A}^2 \over \Gamma_{J/\psi}} &=& { {\rm Br}(J/\psi \rightarrow K^+ K^- \phi) \over \int {{1 \over (2\pi)^3} {1 \over 4 M_{J/\psi}^2} p_{K^+} \widetilde p_{K^-} d M_{\rm inv}(K^- \phi) }}
\\ &=& 0.018\pm0.003~{\rm MeV}^{-1}\, ,
\label{result:tree}
\end{eqnarray}
where the error is taken from Eq.~(\ref{experiment}).

\subsection{Triangle diagram mechanism for the $J/\psi \rightarrow K^+ K^- f_0(980)(a_0(980))$ decays}
\label{sec:triangle}

In the former subsection we studied the $J/\psi \rightarrow K^+ K^- \phi$ decay. In this subsection we show how the \\
$J/\psi \rightarrow K^+ K^- f_0(980)(a_0(980))$ decays can be produced using this input. To do this we look into the triangle diagrams depicted in Fig.~\ref{fig:triangle}. The two mechanisms of Fig.~\ref{fig:triangle}(a) and Fig.~\ref{fig:triangle}(b) are clearly distinguishable. The kinematics of the reaction around the TS point provide for the mechanism of Fig.~\ref{fig:triangle}(a) a momentum $p_{K^+} \approx 1114$~MeV$/c$ and $p_{K^-} \approx 170$~MeV$/c$, and opposite for the mechanism of Fig.~\ref{fig:triangle}(b).
In addition, as we shall see later (see Eq.~(\ref{eq:triangle2})), in one diagram we have ${\vec \epsilon}(J/\psi) \cdot \vec k(K^-)$ and in the other one ${\vec \epsilon}(J/\psi) \cdot \vec k(K^+)$. Upon squaring the sum of the two amplitudes and summing over the $J/\psi$ polarizations, we get the crossed term proportional to $\vec k(K^-) \cdot \vec k(K^+)$, linear in the cosine of the $K^-$ and $K^+$ angle, which will cancel upon angle integrations in the phase space.
Because of this, there is also no interference between these mechanisms. The two mechanisms give the same width and we can study just one of them.

We take the first diagram Fig.~\ref{fig:triangle}(a) and the $J/\psi \rightarrow K^+ K^- f_0(980)$ decay as an example,
and write down its amplitude as:
\begin{eqnarray}
\nonumber - i t &=& - i \mathcal{A} \int{d^4 q \over (2\pi)^4} { i \over (P-q)^2 - m_\phi^2 + i \epsilon } { i \over q^2 - m_{K^-}^2 + i \epsilon }
\\ \nonumber && ~~~~~ \times { i \over (P-q-k)^2 - m_{K^+}^2 + i \epsilon}
\\ \nonumber && ~~~~~ \times {\vec \epsilon}(J/\psi) \cdot {\vec \epsilon}(\phi)~{\vec \epsilon}(\phi) \cdot (2{\vec k} + {\vec q} - {\vec P})
\\ \nonumber && ~~~~~ \times (-i g_V)~(-i g_{f_0,K^+K^-})
\\ \nonumber &=& \mathcal{A} \int{d^4 q \over (2\pi)^4} { 1 \over (P-q)^2 - m_\phi^2 + i \epsilon } { 1 \over q^2 - m_{K^-}^2 + i \epsilon }
\\ \nonumber && ~~~~~ \times { 1 \over (P-q-k)^2 - m_{K^+}^2 + i \epsilon}
\\ && ~~~~~ \times {\vec \epsilon}(J/\psi) \cdot (2{\vec k} + {\vec q} - {\vec P})~g_V~g_{f_0,K^+K^-} \, ,
\label{int:triangle}
\end{eqnarray}
where in the last equation we have summed over the $\phi$ polarization.
The $\phi \rightarrow K^+ K^-$ vertex is obtained from the $P$-wave Lagrangian~\cite{Bando:1987br,Oset:2009vf},
\begin{equation}
\mathcal{L}_{VPP} = - i g_V \langle[P,~\partial_\mu P]~V^\mu\rangle \, , \, g_V = {m_V \over 2f_\pi} \, ,
\end{equation}
where $P$ and $V$ are the ordinary pseudoscalar and vector meson $SU(3)$ matrices, $m_V$ is the vector mass ($m_V \sim 800$~MeV), and $f_\pi$ is the pion decay constant ($f_\pi = 93$~MeV).
This Lagrangian produces a vertex
\begin{equation}
t_{\phi \rightarrow K^+ K^-} = g_V (p^\mu_{K^+} - p^\mu_{K^-}) \epsilon_{\mu}(\phi) \, ,
\end{equation}
and the $\epsilon^0(\phi)$ component can be neglected as shown in Appendix A of Ref.~\cite{Sakai:2017hpg}, since the three-momentum of the $\phi$ is very small compared to its mass for $\phi$ on-shell in that diagram.
The $f_0(980) \rightarrow K^+ K^-$ and $a_0(980) \rightarrow K^+ K^-$ vertices are obtained from the chiral unitary approach of Ref.~\cite{Oller:1997ti}
with $g_{f_0,K^+K^-} = 2567$~MeV and $g_{a_0,K^+K^-} = 3875$~MeV. Note that these two vertices will be more carefully examined in the next subsection.

Then we follow Refs.~\cite{Bayar:2016ftu,Aceti:2015zva} and perform analytically the $q^0$ integration in Eq.~(\ref{int:triangle}) in the $K^- f_0(980)$ rest frame, with the result
\begin{eqnarray}
\nonumber t &=& \mathcal{A} \int{d^3 q \over (2\pi)^3}
{ 1 \over 8 \omega_{K^+} \omega_{K^-} \omega_{\phi} }
{ 1 \over k^0 - \omega_{K^+} - \omega_{\phi} + i \Gamma_\phi/2 }
\\ \nonumber && ~~~~~ \times { 1 \over M_{\rm inv}(K^- f_0) + \omega_{K^-} + \omega_{K^+} - k^0}
\\ \nonumber && ~~~~~ \times { 1 \over M_{\rm inv}(K^- f_0) - \omega_{K^-} - \omega_{K^+} - k^0 + i \epsilon}
\\ \nonumber && ~~~~~ \times {1 \over M_{\rm inv}(K^- f_0) - \omega_{\phi} - \omega_{K^-} + i \Gamma_\phi/2 }
\\ \nonumber && ~~~~~ \times \Bigg( 2 M_{\rm inv}(K^- f_0)\omega_{K^-} + 2 k^0 \omega_{K^+}
\\ \nonumber && ~~~~~~~~~~~~~~~ - 2(\omega_{K^-} + \omega_{K^+})(\omega_{K^-} + \omega_{K^+} + \omega_{\phi}) \Bigg)
\\ && ~~~~~ \times g_V g_{f_0,K^+K^-}~{\vec \epsilon}(J/\psi) \cdot (2{\vec k} + {\vec q}) \, ,
\label{eq:triangle}
\end{eqnarray}
where
\begin{eqnarray}
\nonumber \omega_{K^-} &=& \sqrt{m_{K^-}^2 + \vec q^2 } \, ,
\\ \nonumber \omega_{K^+} &=& \sqrt{m_{K^+}^2 + (\vec q + \vec k)^2 } \, ,
\\ \nonumber \omega_{\phi} &=& \sqrt{m_{\phi}^2 + \vec q^2 } \, ,
\\ \nonumber k^0 &=& { M^2_{\rm inv}(K^- f_0) + m_{K^-}^2 - m_{f_0}^2 \over 2 M_{\rm inv}(K^- f_0) } \, ,
\\ k = |\vec k| &=& {\lambda^{1/2} (M^2_{\rm inv}(K^- f_0) , m_{K^-}^2 , m_{f_0}^2 ) \over 2 M_{\rm inv}(K^- f_0) } \, .
\end{eqnarray}
In addition, since $\vec k$ is the only momentum not integrated in Eq.~(\ref{eq:triangle}), we can replace
\begin{eqnarray}
\int d^3 q~\vec q \longrightarrow \vec k \int d^3 q~{\vec q \cdot \vec k \over \vec k^2} \, .
\end{eqnarray}
Then the amplitude $t$ of Eq.~(\ref{eq:triangle}) can be rewritten as
\begin{eqnarray}
t &=& \mathcal{A}~g_V~g_{f_0,K^+K^-}~{\vec \epsilon}(J/\psi) \cdot \vec k~t_T \, ,
\label{eq:triangle2}
\end{eqnarray}
where
\begin{eqnarray}
\nonumber t_T &=& \int{d^3 q \over (2\pi)^3}
{ 1 \over 8 \omega_{K^+} \omega_{K^-} \omega_{\phi} }
{ 1 \over k^0 - \omega_{K^+} - \omega_{\phi} + i \Gamma_\phi/2 }
\\ \nonumber && ~~~~~ \times { 1 \over M_{\rm inv}(K^- f_0) + \omega_{K^-} + \omega_{K^+} - k^0}
\\ \nonumber && ~~~~~ \times { 1 \over M_{\rm inv}(K^- f_0) - \omega_{K^-} - \omega_{K^+} - k^0 + i \epsilon}
\\ \nonumber && ~~~~~ \times {1 \over M_{\rm inv}(K^- f_0) - \omega_{\phi} - \omega_{K^-} + i \Gamma_\phi/2 }
\\ \nonumber && ~~~~~ \times \Bigg( 2 M_{\rm inv}(K^- f_0)\omega_{K^-} + 2 k^0 \omega_{K^+}
\\ \nonumber && ~~~~~~~~~~~~~~~ - 2(\omega_{K^-} + \omega_{K^+})(\omega_{K^-} + \omega_{K^+} + \omega_{\phi}) \Bigg)
\\ && ~~~~~ \times \left(2 + {{\vec q \cdot \vec k} \over \vec k^2}\right) \, .
\label{eq:tt}
\end{eqnarray}
As in Ref.~\cite{Bayar:2016ftu}, the above integration is regularized with the factor $\theta(q_{\rm max} - |\vec q^{\, *}|)$, where $\vec q^{\, *}$ is the momentum of the $K^-$ in the rest frame of $f_0(980)$,
with $q_{\rm max} = 600$~MeV as it is needed in the chiral unitary approach that reproduces the $f_0(980)$~\cite{Liang:2014tia,Xie:2014tma}. After summing over polarizations in Eq.~(\ref{eq:triangle2}), we obtain
\begin{eqnarray}
\nonumber && \overline{\sum} \sum |t|^2
\\ \nonumber &=& \sum \mathcal{A}^2~g_V^2~g^2_{f_0,K^+K^-}~|t_T|^2~{1\over3}~{\vec \epsilon}(J/\psi) \cdot \vec k~{\vec \epsilon}(J/\psi) \cdot \vec k
\\ &=& \mathcal{A}^2~g_V^2~g^2_{f_0,K^+K^-}~|t_T|^2~{1\over3}~\left|{\vec k}\right|^2 \, .
\label{result:trianglesum}
\end{eqnarray}

Now we can write the $K^- f_0(980)$ invariant mass distribution of the $J/\psi \rightarrow K^+ K^- f_0(980)$ decay as
\begin{equation}
{d \Gamma_{J/\psi \rightarrow K^+ K^- f_0(980)} \over d M_{\rm inv}(K^- f_0)} = {1 \over (2\pi)^3} {1 \over 4 M_{J/\psi}^2} p^\prime_{K^+} \widetilde p^{\, \prime}_{K^-} \overline{\sum} \sum |t|^2 \, .
\end{equation}
where $p^\prime_{K^+}$ is the momentum of the $K^+$ in the $J/\psi$ rest frame, and $\widetilde p^{\, \prime}_{K^-} = k$ is the momentum of the $K^-$ in the $K^- f_0(980)$ rest frame:
\begin{eqnarray}
\nonumber \ p^\prime_{K^+} &=& {\lambda^{1/2}(M_{J/\psi}^2, m_{K^+}^2, M^2_{\rm inv}(K^- f_0)) \over 2 M_{J/\psi} }  \, ,
\\
\widetilde p^{\, \prime}_{K^-} &=& k = {\lambda^{1/2} (M^2_{\rm inv}(K^- f_0) , m_{K^-}^2 , m_{f_0}^2 ) \over 2 M_{\rm inv}(K^- f_0) } \, .
\end{eqnarray}
Recalling Eq.~(\ref{result:tree}) and Eq.~(\ref{result:trianglesum}), we obtain the differential branching ratio of the $J/\psi \rightarrow K^+ K^- f_0(980)$ decay to be
\begin{eqnarray}
&& {1\over\Gamma_{J/\psi}}{d \Gamma_{J/\psi \rightarrow K^+ K^- f_0(980)} \over d M_{\rm inv}(K^- f_0)}
\label{eq:trianglegamma}
\\ \nonumber &=& {\mathcal{A}^2\over\Gamma_{J/\psi}}{1 \over (2\pi)^3} {1 \over 4 M_{J/\psi}^2}~{1\over3} p^\prime_{K^+} \widetilde p^{\,\prime3}_{K^-}~g_V^2 g^2_{f_0,K^+K^-}~|t_T|^2 \, .
\end{eqnarray}
The case for $a_0(980)$ production is identical replacing $g_{f_0,K^+K^-}$ by $g_{a_0,K^+K^-}$.



\subsection{The $J/\psi \rightarrow K^+ K^- f_0(980)\rightarrow K^+ K^- \pi^+\pi^-$ and $J/\psi \rightarrow K^+ K^- a_0(980) \rightarrow K^+ K^- \pi^0 \eta$ reactions}
\label{sec:fourfinal}

%
\begin{figure*}[hbt]
\begin{center}
\subfigure[]{
\includegraphics[width=0.4\textwidth]{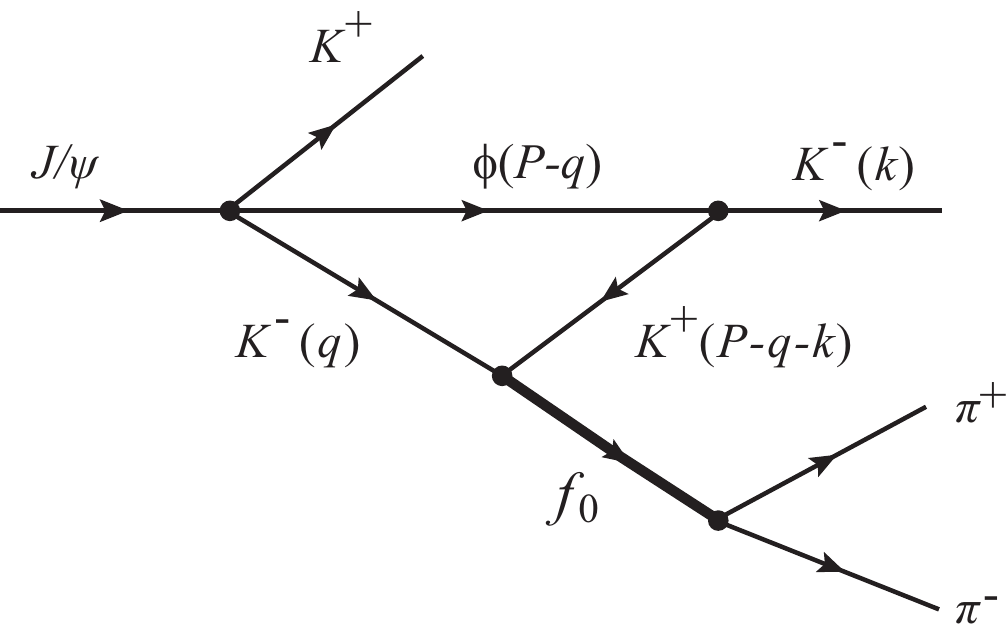}}
~~~~~~~~~~
\subfigure[]{
\includegraphics[width=0.4\textwidth]{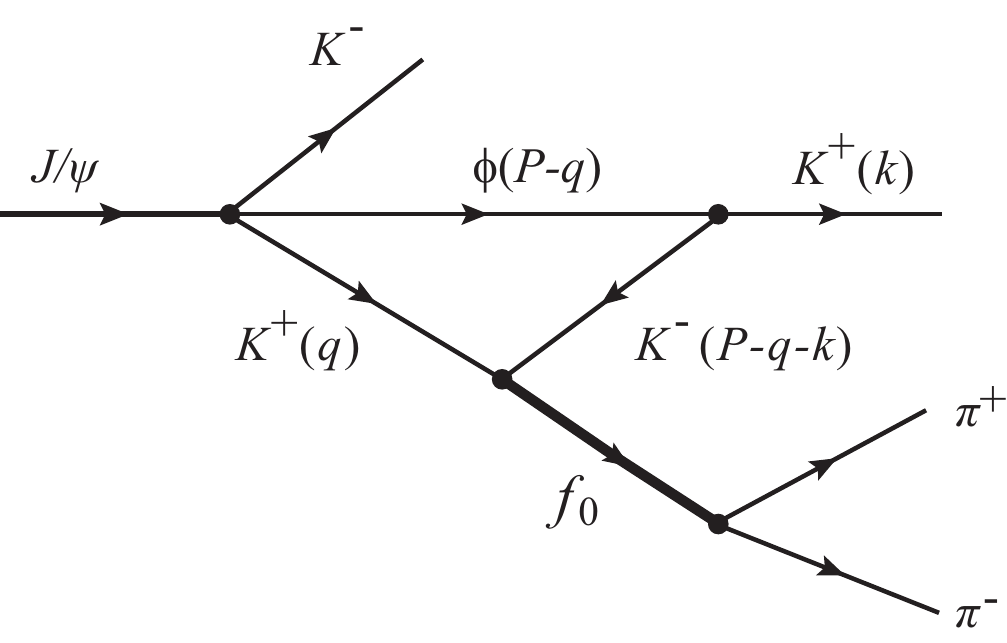}}
\\
\subfigure[]{
\includegraphics[width=0.4\textwidth]{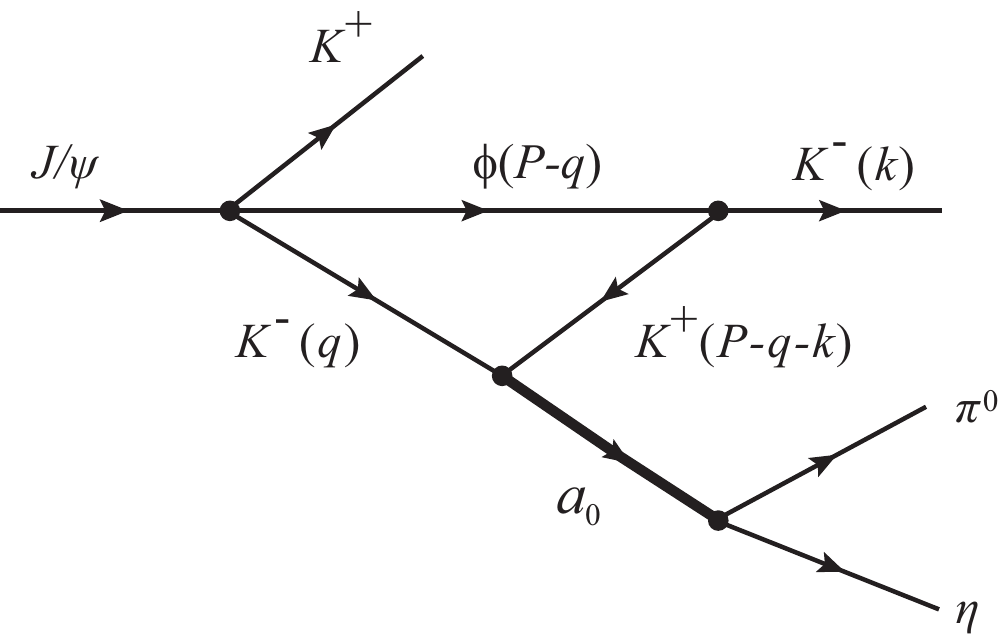}}
~~~~~~~~~~
\subfigure[]{
\includegraphics[width=0.4\textwidth]{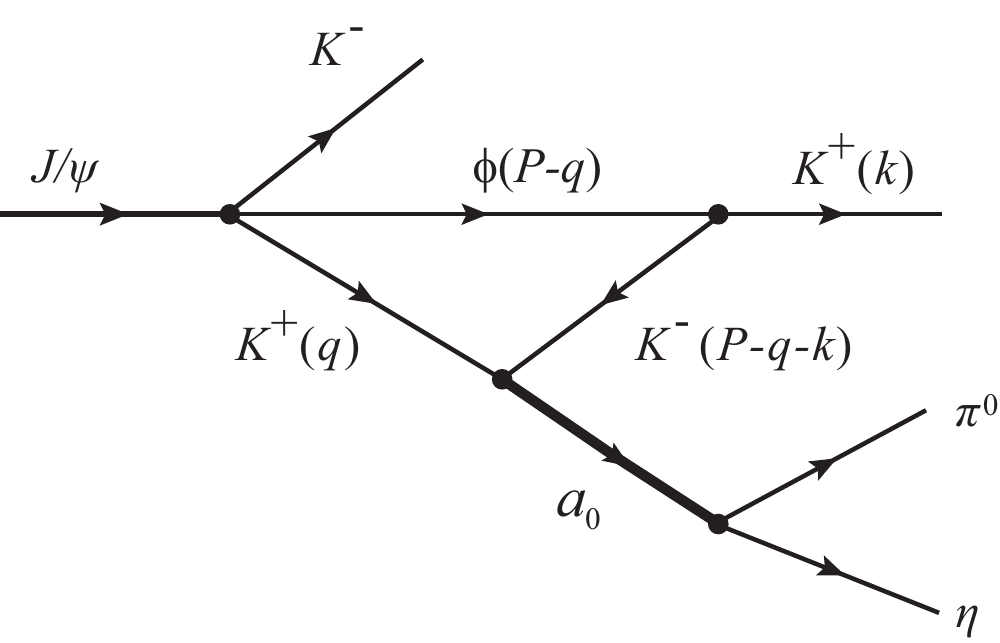}}
\caption{Triangle diagrams for the (a, b) $J/\psi \rightarrow K^+ K^- f_0(980) \rightarrow K^+ K^- \pi^+\pi^-$ and (c, d) $J/\psi \rightarrow K^+ K^- a_0(980) \rightarrow K^+ K^- \pi^0 \eta$ reactions.}
\label{fig:fourfinal}
\end{center}
\end{figure*}
%

In the former subsection we studied the triangle diagram mechanism for the $J/\psi \rightarrow K^+ K^- f_0(980)(a_0(980))$ decays. In this subsection we further consider that the $f_0(980)$ and $a_0(980)$ will be seen in the $\pi^+\pi^-$ and $\pi^0\eta$ mass distribution, respectively, as depicted in Fig.~\ref{fig:fourfinal}. Take the first diagram Fig.~\ref{fig:fourfinal}(a) as an example, the $J/\psi$ first decays into $K^+ K^- \phi$, next the $\phi$ decays into $K^+ K^-$, then the $K^-$ and $K^+$ merge to give the $f_0(980)$, and finally the $f_0(980)$ decays into $\pi^+ \pi^-$.
We can write down its amplitude as:
\begin{eqnarray}
\nonumber - i t^\prime &=& - i \mathcal{A} \int{d^4 q \over (2\pi)^4} { i \over (P-q)^2 - m_\phi^2 + i \epsilon }{ i \over q^2 - m_{K^-}^2 + i \epsilon }
\\ \nonumber && ~~~~~ \times { i \over (P-q-k)^2 - m_{K^+}^2 + i \epsilon}
\\ \nonumber && ~~~~~ \times {\vec \epsilon}(J/\psi) \cdot {\vec \epsilon}(\phi)~{\vec \epsilon}(\phi) \cdot (2{\vec k} + {\vec q} - {\vec P})
\\ \nonumber && ~~~~~ \times (-i g_V)~(-i t_{K^+K^-,\pi^+\pi^-})
\\ \nonumber &=& \mathcal{A} \int{d^4 q \over (2\pi)^4} { 1 \over (P-q)^2 - m_\phi^2 + i \epsilon } { 1 \over q^2 - m_{K^-}^2 + i \epsilon }
\\ \nonumber && ~~~~~ \times { 1 \over (P-q-k)^2 - m_{K^+}^2 + i \epsilon}
\\ \nonumber && ~~~~~ \times {\vec \epsilon}(J/\psi) \cdot (2{\vec k} + {\vec q} - {\vec P})~g_V~t_{K^+K^-,\pi^+\pi^-} \, .
\\ \label{int:fourfinal}
\end{eqnarray}
This amplitude $t^\prime$ is very similar to $t$ given in Eq.~(\ref{int:triangle}) in the former subsection, just with $g_{f_0,K^+K^-}$ replaced by the transition amplitude $t_{K^+K^-,\pi^+\pi^-}$. We can follow the same procedure to simplify it to be
\begin{eqnarray}
t^\prime &=& \mathcal{A}~g_V~t_{K^+K^-,\pi^+\pi^-}~{\vec \epsilon}(J/\psi) \cdot \vec k^{\prime}~t^\prime_T \, ,
\label{eq:fourfinal}
\end{eqnarray}
where $t^\prime_T$ is also very similar to $t_T$ given in Eq.~(\ref{eq:tt}), just with the following replacements:
\begin{eqnarray}
k^0 &\rightarrow& k^{\prime0} = { M^2_{\rm inv}(K^- f_0) + m_{K^-}^2 - M^2_{\rm inv}(\pi^+ \pi^-) \over 2 M_{\rm inv}(K^- f_0) } \, ,
\label{eq:k0}
\\ k &\rightarrow& k^\prime = {\lambda^{1/2} (M^2_{\rm inv}(K^- f_0) , m_{K^-}^2 , M^2_{\rm inv}(\pi^+ \pi^-) ) \over 2 M_{\rm inv}(K^- f_0) } \, .
\label{eq:k}
\end{eqnarray}
The $K^+ K^- \rightarrow \pi^+ \pi^-$ and $K^+ K^- \rightarrow \pi^0 \eta$ scattering has been studied in detail in Refs.~\cite{Liang:2014tia,Xie:2014tma} within the chiral unitary approach, where altogether six channels were taken into account, including $\pi^+\pi^-$, $\pi^0\pi^0$, $K^+K^-$, $K^0 \bar K^0$, $\eta \eta$, and $\pi^0\eta$. In the present study we use this as input, and we shall see simultaneously both the $f_0(980)$ (with $I = 0$) and $a_0(980)$ (with $I = 1$)
productions.

Now we can write down the double differential mass distribution for the $J/\psi \rightarrow K^+ K^- f_0(980)\rightarrow K^+ K^- \pi^+\pi^-$ reaction, as a function of $M_{\rm inv}(K^- f_0)$ and $M_{\rm inv}(\pi^+\pi^-)$~\cite{Pavao:2017kcr}:
\begin{eqnarray}
\nonumber && {d^2 \Gamma_{J/\psi \rightarrow K^+ K^- f_0(980)\rightarrow K^+ K^- \pi^+\pi^-} \over d M_{\rm inv}(K^- f_0)  d M_{\rm inv}(\pi^+ \pi^-)}
\\ &=& {1 \over (2\pi)^5} {1 \over 4 M_{J/\psi}^2} p^{\prime\prime}_{K^+} \widetilde p^{\,\prime\prime}_{K^-} \widetilde p^{\,\prime\prime}_{\pi^+} \overline{\sum} \sum |t^\prime|^2 \, ,
\end{eqnarray}
where $p^{\prime\prime}_{K^+}$ is the momentum of the $K^+$ in the $J/\psi$ rest frame, $\widetilde p^{\, \prime\prime}_{K^-} = k^\prime$ is the momentum of the $K^-$ in the $K^- f_0(980)$ rest frame, and $\widetilde p^{\, \prime\prime}_{\pi^+}$ is the momentum of the $\pi^+$ in the $\pi^+ \pi^-$ rest frame:
\begin{eqnarray}
\nonumber p^{\prime\prime}_{K^+} &=& {\lambda^{1/2}(M_{J/\psi}^2, m_{K^+}^2, M^2_{\rm inv}(K^- f_0)) \over 2 M_{J/\psi} }  \, ,
\\
\nonumber \widetilde p^{\, \prime\prime}_{K^-} &=& k^\prime = {\lambda^{1/2} (M^2_{\rm inv}(K^- f_0) , m_{K^-}^2 , M^2_{\rm inv}(\pi^+ \pi^-) ) \over 2 M_{\rm inv}(K^- f_0) } \, ,
\\
\widetilde p^{\, \prime\prime}_{\pi^+} &=& {\lambda^{1/2} (M^2_{\rm inv}(\pi^+ \pi^-) , m_{\pi^+}^2 , m_{\pi^-}^2 ) \over 2 M_{\rm inv}(\pi^+ \pi^-) } \, .
\end{eqnarray}
Recalling Eq.~(\ref{result:tree}) and Eq.~(\ref{eq:fourfinal}), we obtain the double differential branching ratio of the $J/\psi \rightarrow K^+ K^- f_0(980)\rightarrow K^+ K^- \pi^+\pi^-$ reaction to be
\begin{eqnarray}
\nonumber && {1\over\Gamma_{J/\psi}}{d^2 \Gamma_{J/\psi \rightarrow K^+ K^- f_0(980)\rightarrow K^+ K^- \pi^+\pi^-} \over d M_{\rm inv}(K^- f_0)  d M_{\rm inv}(\pi^+ \pi^-)}
\\ \nonumber &=& {\mathcal{A}^2\over\Gamma_{J/\psi}} {1 \over (2\pi)^5} {1 \over 4 M_{J/\psi}^2}~{1\over3} p^{\prime\prime}_{K^+} \widetilde p^{\,\prime\prime3}_{K^-} \widetilde p^{\,\prime\prime}_{\pi^+}
\\ & \times& g_V^2~|t_{K^+K^- ,\pi^+ \pi^-}|^2~|t_T^\prime|^2 \, .
\label{eq:doublegamma}
\end{eqnarray}
With trivial changes, replacing $\pi^+\pi^-$ by $\pi^0 \eta$, we get the corresponding expressions for $\pi^0 \eta$ production.

One should note that a different picture of these resonances, like compact $\bar q q$ or tetraquarks, which happened to provide the same couplings to $K \bar K$, $\pi \pi$, $\pi \eta$, would produce the same results that we have reported here. The difference of the two pictures stems from the fact that for the dynamically generated resonances that come from the interaction of the $K \bar K$ and coupled channels, it is the interaction of the $K \bar K$ primarily produced that produces the $f_0$ and $a_0$ in the final state. A compact quark state, $Q$, necessarily has overlap with $J/\psi$ and $K^+K^-$ in a direct $J/\psi \to K^+ K^- Q$ reaction. The $J/\psi \rightarrow K^+ K^- f_0(a_0)$ production contains then two mechanisms, the direct one, that one can envisage important for a compact object, and the triangle mechanism. For a dynamically generated resonance the triangle mechanism is the only production mode. Hence, there would be different predictions for the production rate of $f_0(a_0)$ in these pictures, but lacking the predictions from the compact picture, one cannot go any further. There are not many of such reactions where evaluations are done with both pictures, but some exist, like the $B_s \to J/\psi f_0$ reaction~\cite{Aaij:2011fx}, where a compact picture is proposed in Ref.~\cite{Stone:2013eaa} and a more detailed description of the data in line with the molecular picture is done in Ref.~\cite{Liang:2014tia} (see also related method in Ref.~\cite{Daub:2015xja}). Other cases are discussed in Ref.~\cite{Oset:2016lyh}.

Ultimately, it is the consistent and systematic description of experimental facts what gives weight to a particular picture, and the molecular picture has passed a great deal of these tests~\cite{Oset:2016lyh,Pelaez:2015qba}. The agreement with experiment of the present study should be considered within this perspective. In any case the main aim of the present work is to point out the presence of the TS in this reaction, with the double purpose of finding experimental cases, where TS are manifested, and anticipating a warning for not confusing the peaks predicted, when observed, with new resonances.

\section{Results}
\label{sec:results}

\begin{figure*}[hbt]
\begin{center}
\subfigure[]{
\includegraphics[width=0.3\textwidth]{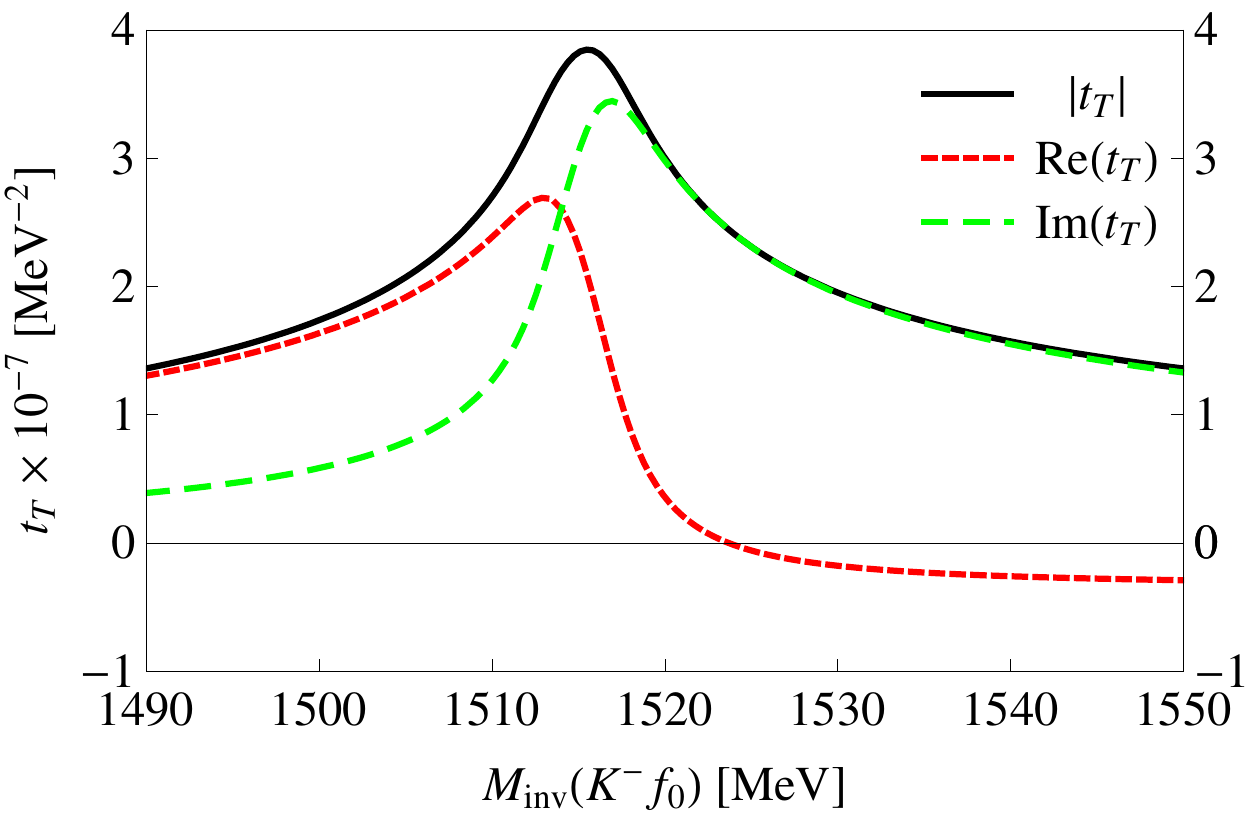}}
\subfigure[]{
\includegraphics[width=0.3\textwidth]{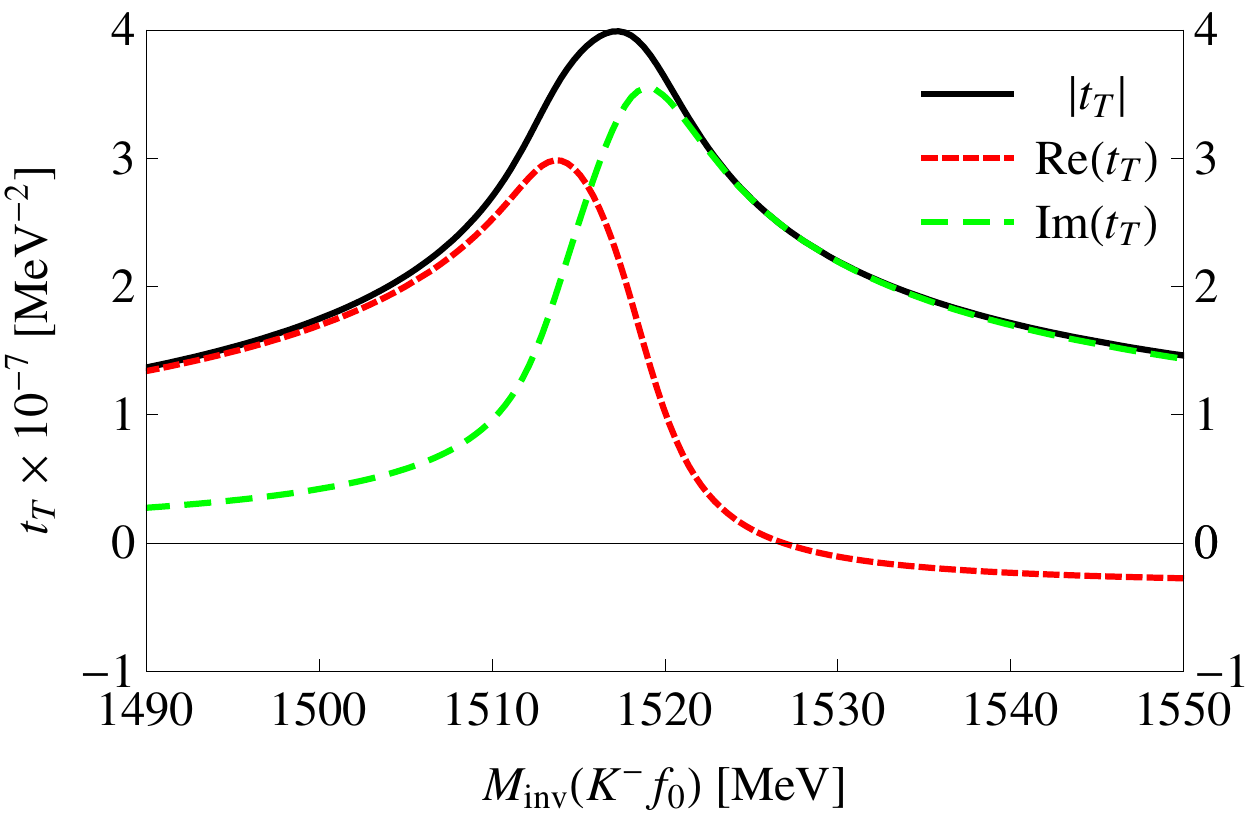}}
\subfigure[]{
\includegraphics[width=0.3\textwidth]{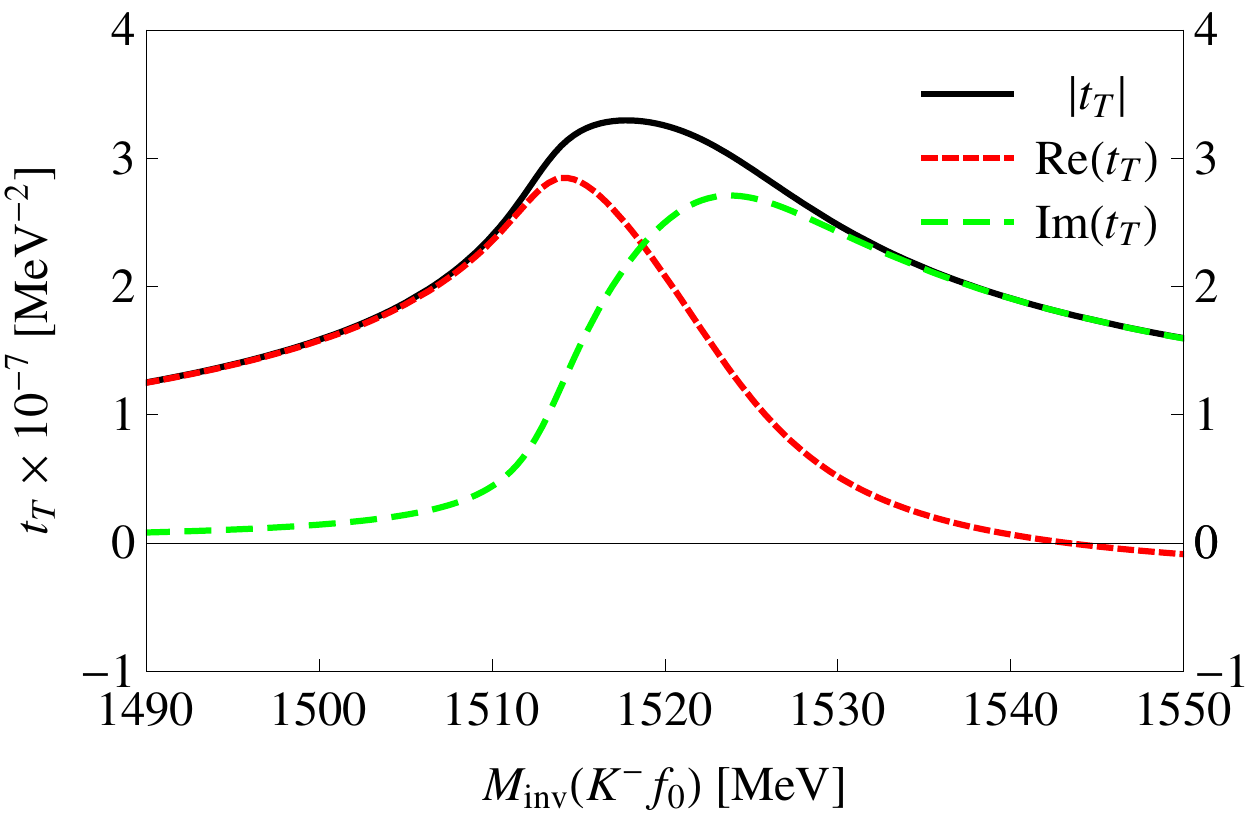}}
\subfigure[]{
\includegraphics[width=0.3\textwidth]{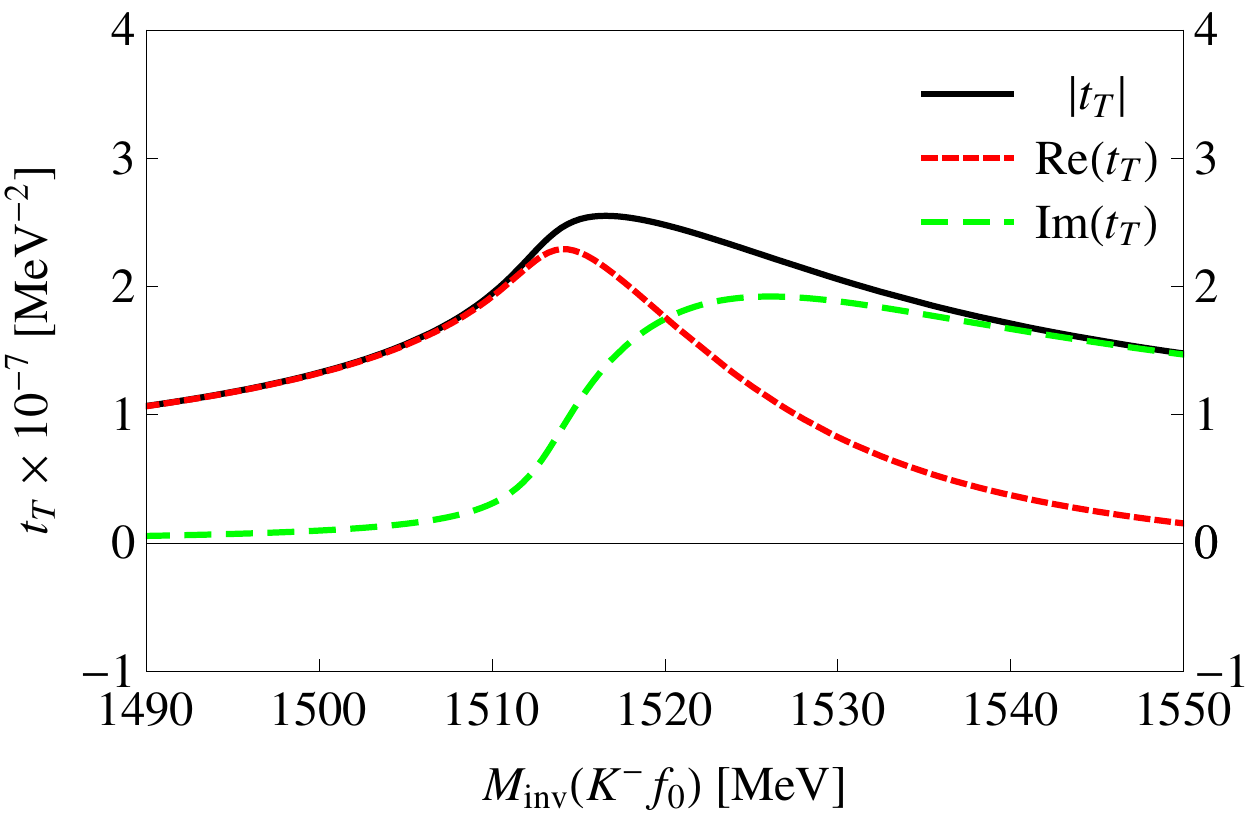}}
\subfigure[]{
\includegraphics[width=0.3\textwidth]{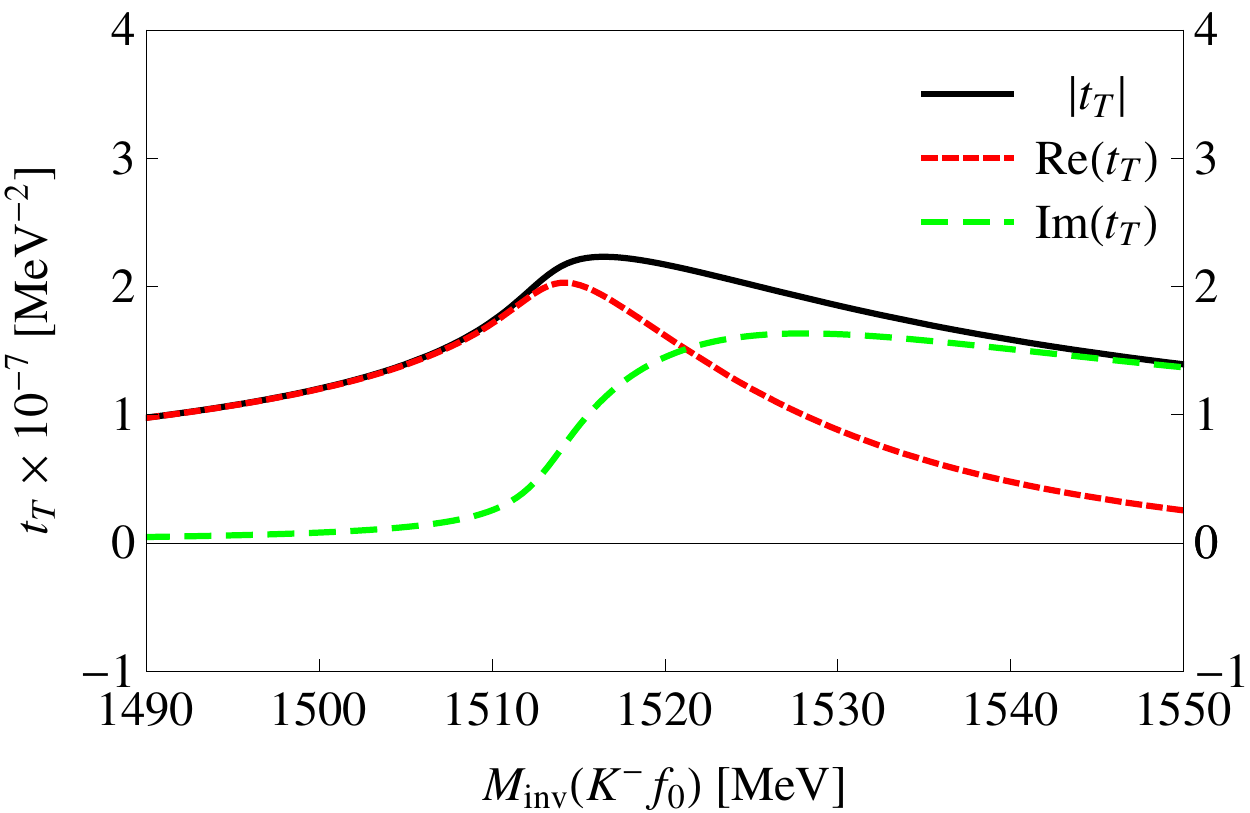}}
\subfigure[]{
\includegraphics[width=0.3\textwidth]{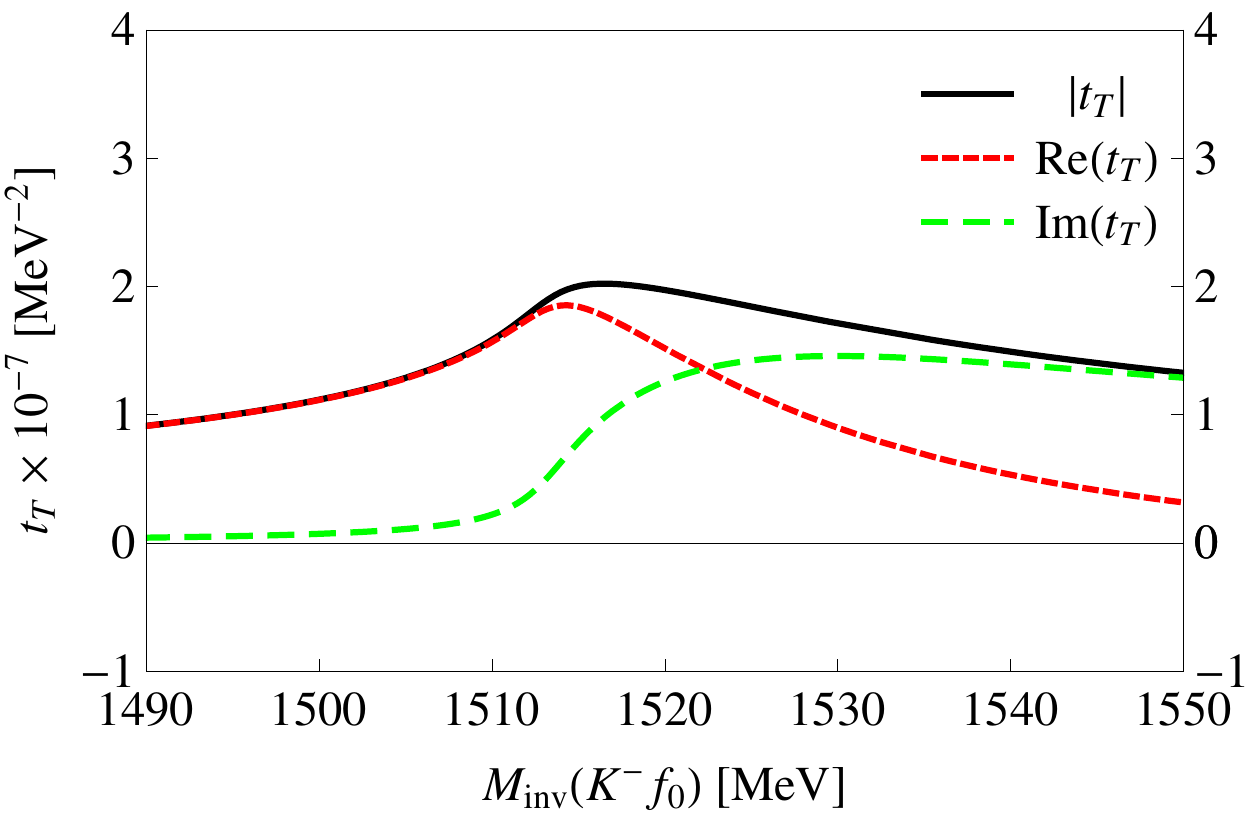}}
\caption{Triangle amplitude $t_T$, defined in Eq.~(\ref{eq:tt}), as a function of $M_{\rm inv}(K^- f_0/K^- a_0)$ for (a) $m_{f_0(a_0)} = 989$~MeV, (b) $m_{f_0(a_0)} = 988$~MeV, (c) $m_{f_0(a_0)} = 987$~MeV, (d) $m_{f_0(a_0)} = 985$~MeV, (e) $m_{f_0(a_0)} = 983$~MeV, and (f) $m_{f_0(a_0)} = 981$~MeV. $|t_T|$, Re$(t_T)$, and Im$(t_T)$ are plotted using the black, red, and green curves, respectively.
}
\label{fig:tt}
\end{center}
\end{figure*}

Firstly, we show our results for the \\
$J/\psi \rightarrow K^+ K^- f_0(980)(a_0(980))$ decays, which were previously studied in Sec.~\ref{sec:triangle}. Let us begin by showing in Fig.~\ref{fig:tt} the contribution of the triangle loop defined in Eq.~(\ref{eq:tt}).
The TS condition of Eq.~(\ref{con:triangle}) requires all $K^+ K^- \phi$ intermediate particles to be on shell. This forces $m_{f_0(a_0)} > m_{K^+} + m_{K^-}$. On the other hand, if we go to energies a bit bigger than that, Eq.~(\ref{con:triangle}) is no longer fulfilled. There is hence a very narrow window of $f_0(a_0)$ masses where the TS condition is exactly fulfilled, {\it i.e.}, from 987 MeV to 993 MeV. In view of this we plot in Fig.~\ref{fig:tt} the real, imaginary parts and modulus of $t_T$ of Eq.~(\ref{eq:tt}) for different masses of $f_0(a_0)$. The magnitude depends on the $f_0(a_0)$ mass, independent on whether we have $f_0$ or $a_0$, since the different couplings to $K^+K^-$ have been factorized out of the integral of $t_T$. We show the results for six different masses. The first two are inside the window of energies where the TS appears, the other four are outside. We observe a neat peak in the first two cases, which gets broader gradually as we depart from the TS window.
Note that the peak of the imaginary part is related to the triangle singularity, while the one of the real part is related to the $K^- \phi$ threshold, as discussed in Refs.~\cite{Sakai:2017hpg,Dai:2018hqb}.

\begin{figure*}[hbt]
\begin{center}
\subfigure[]{
\includegraphics[width=0.3\textwidth]{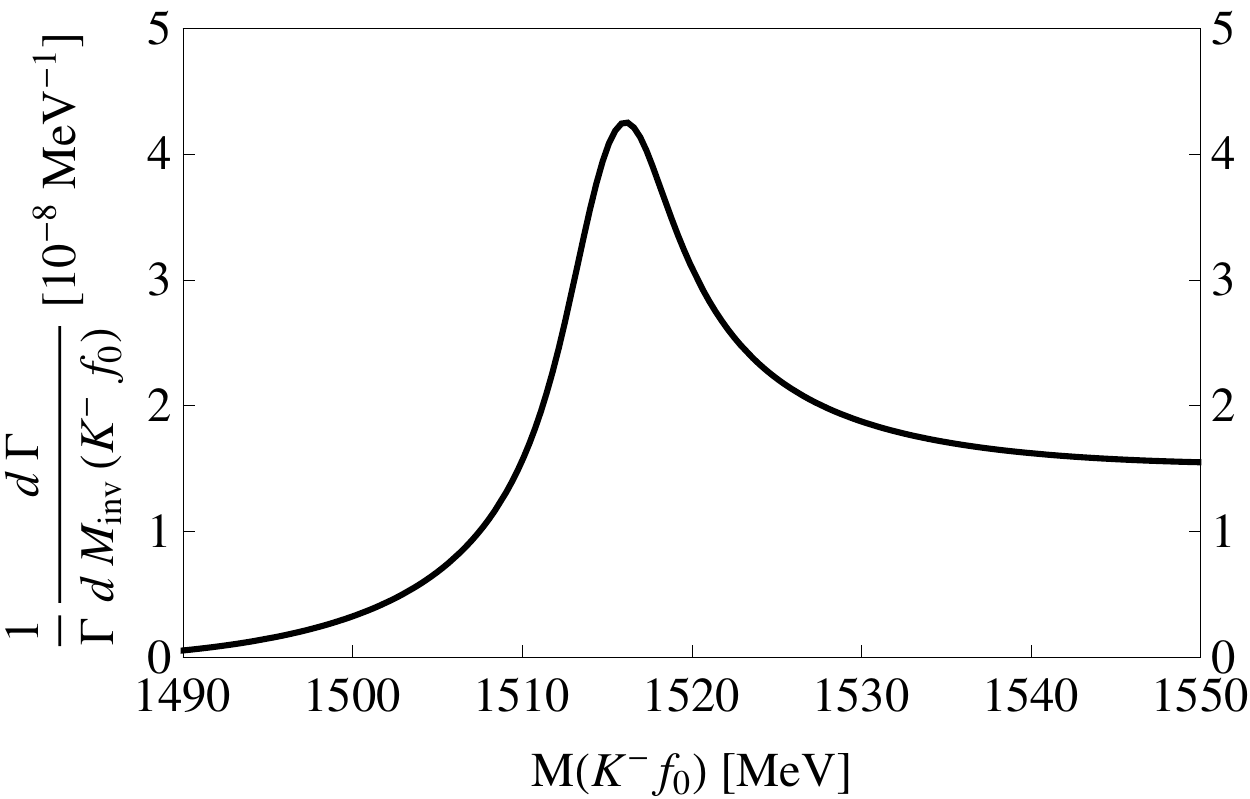}}
\subfigure[]{
\includegraphics[width=0.3\textwidth]{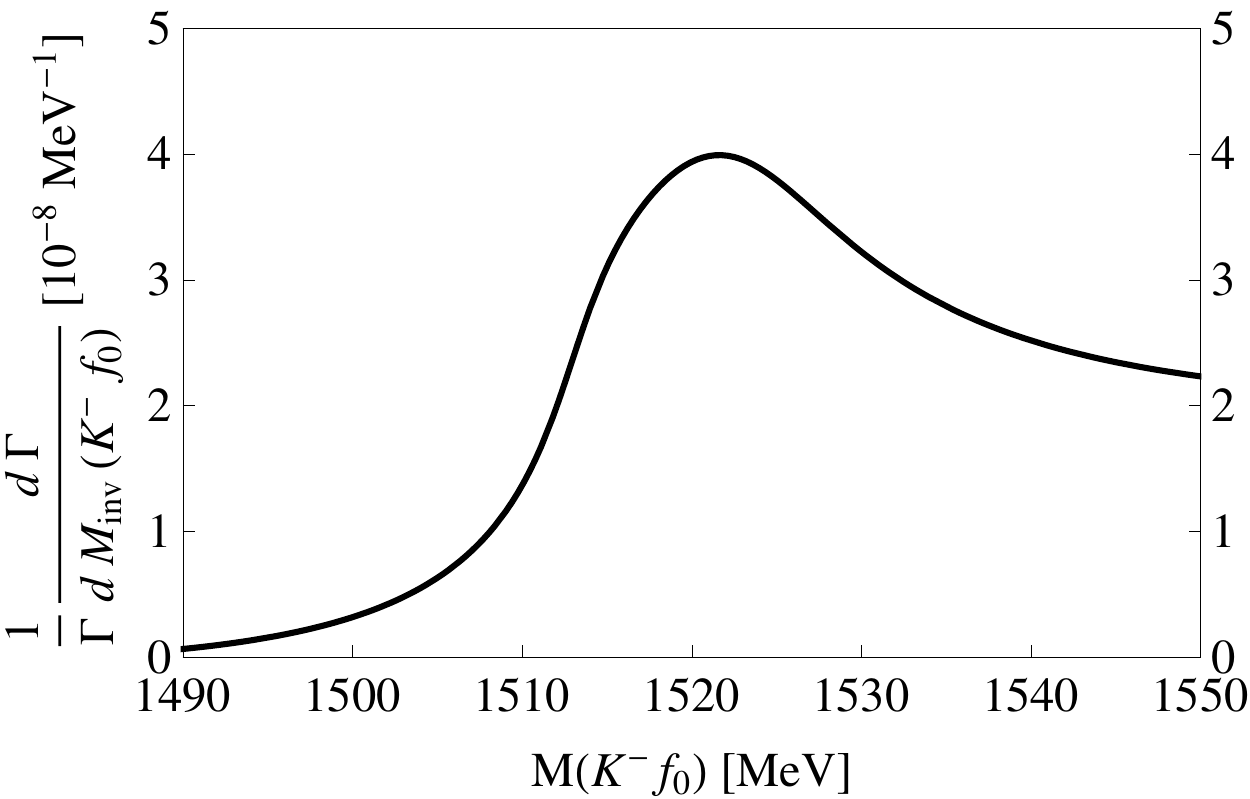}}
\subfigure[]{
\includegraphics[width=0.3\textwidth]{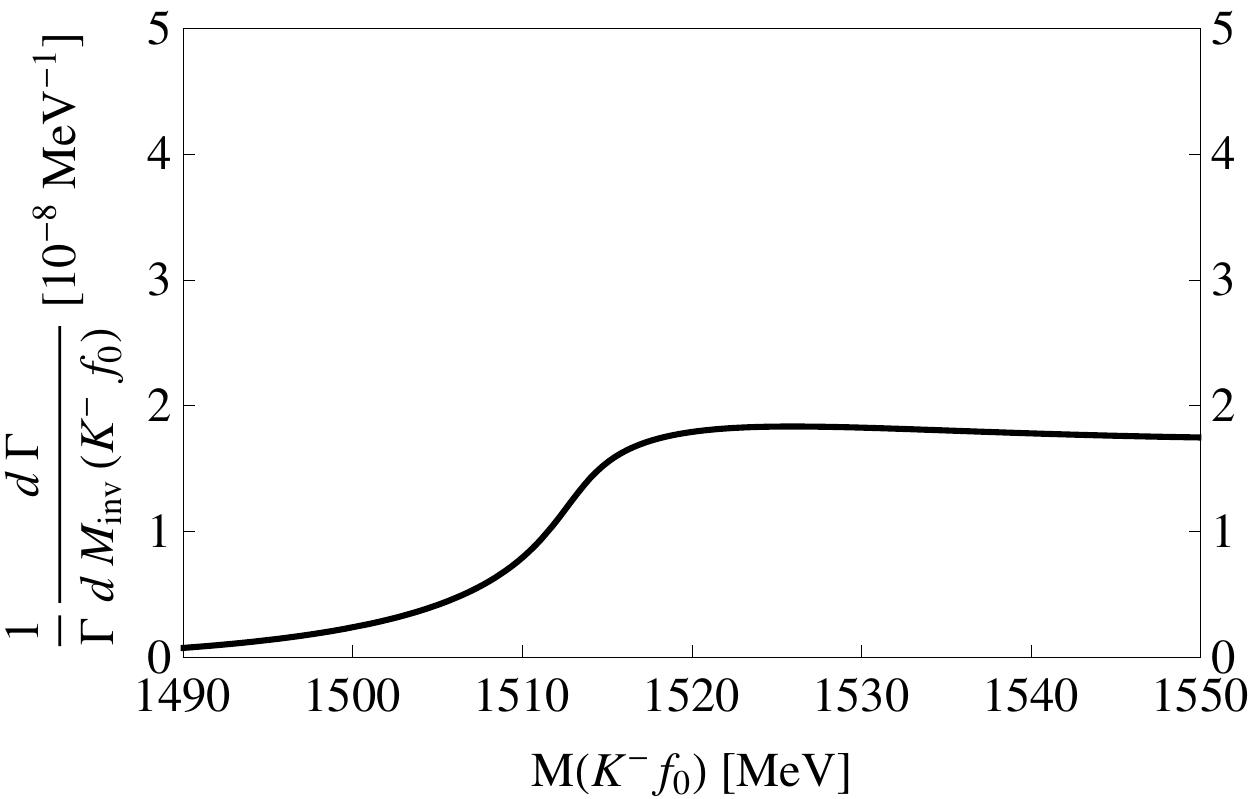}}
\subfigure[]{
\includegraphics[width=0.3\textwidth]{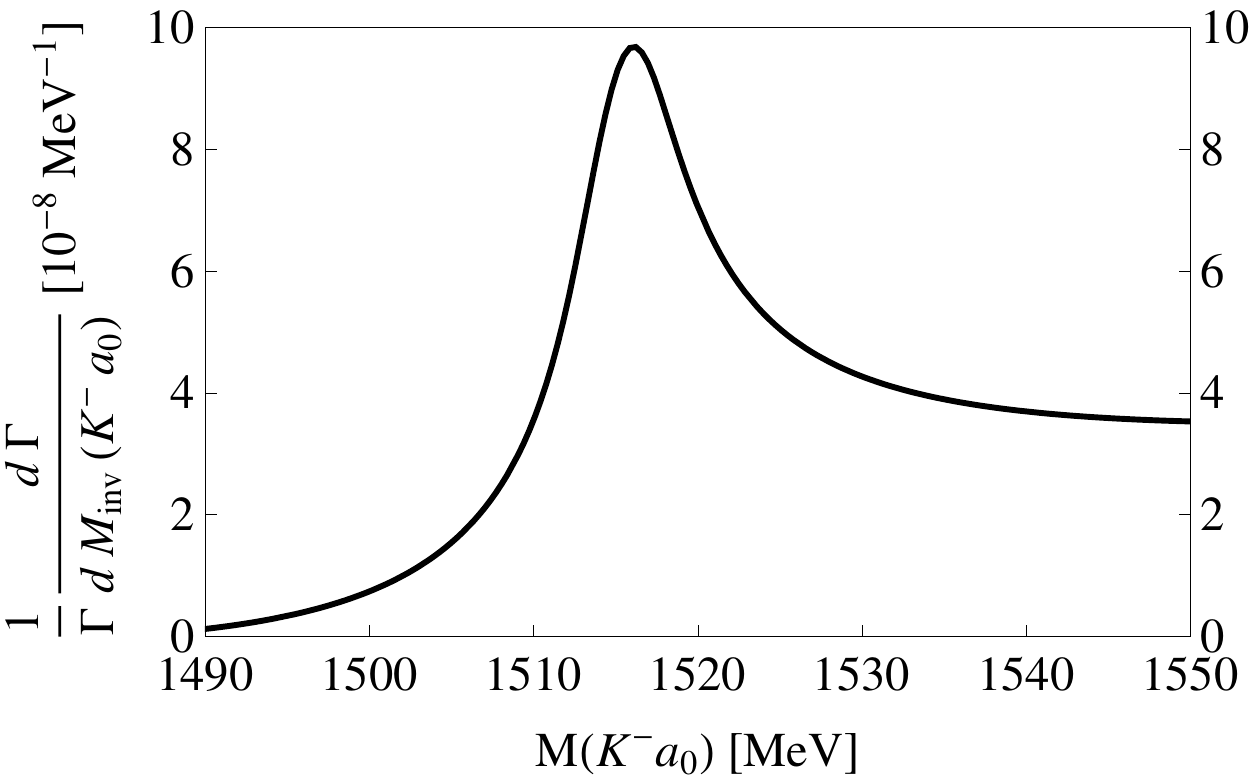}}
\subfigure[]{
\includegraphics[width=0.3\textwidth]{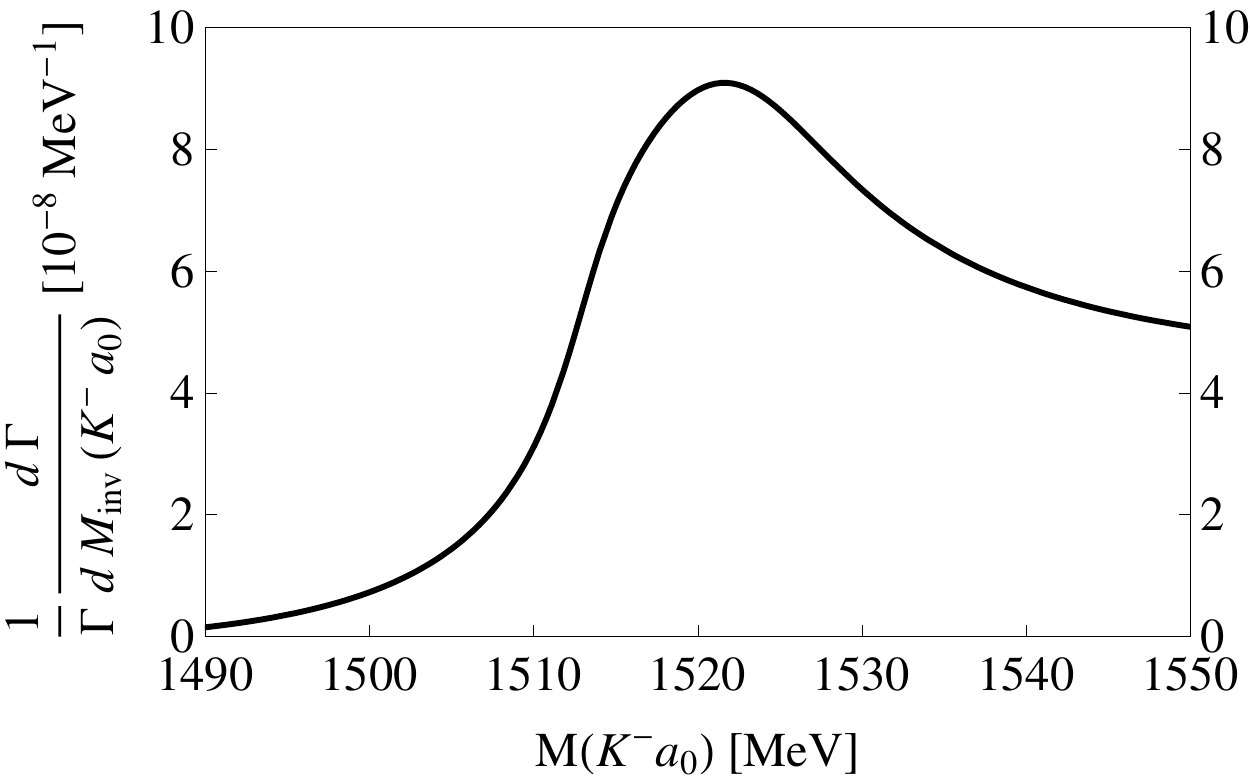}}
\subfigure[]{
\includegraphics[width=0.3\textwidth]{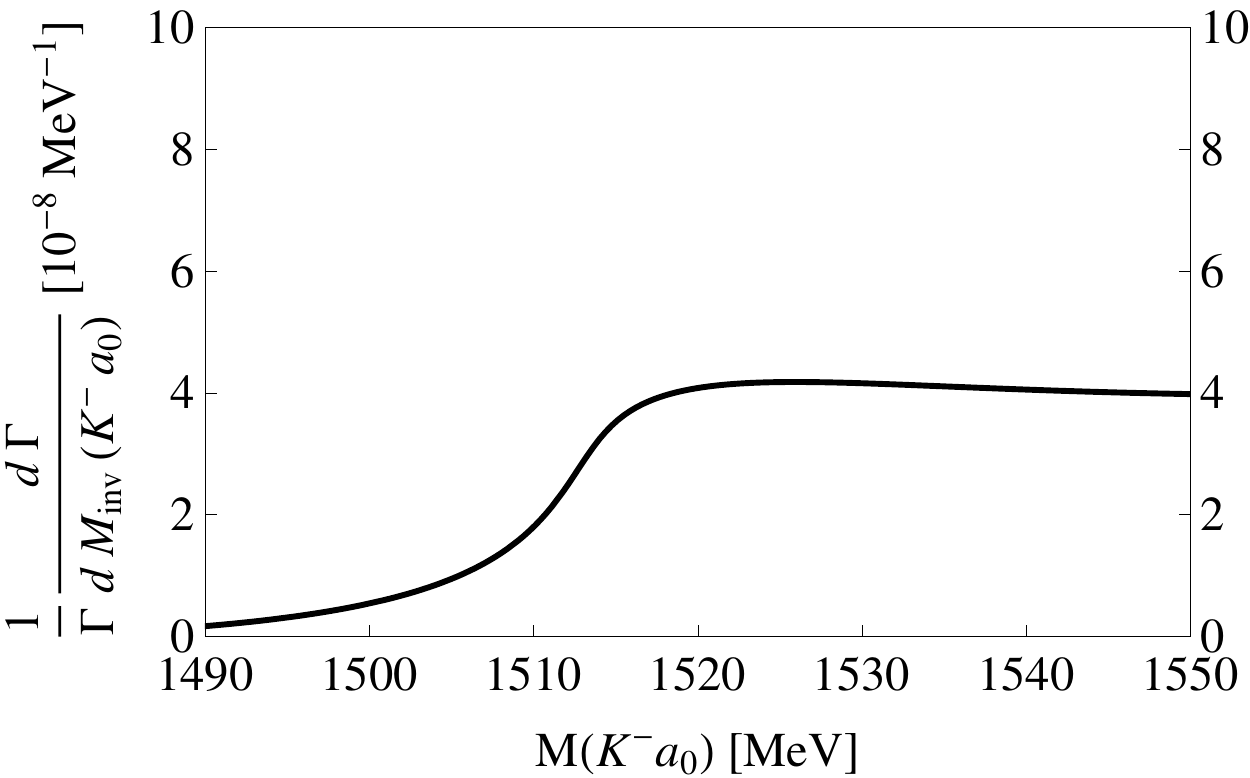}}
\caption{The differential branching ratio ${1\over\Gamma_{J/\psi}}{d \Gamma_{J/\psi \rightarrow K^+ K^- f_0(a_0)} \over d M_{\rm inv}(K^- f_0/K^- a_0)}$, defined in Eq.~(\ref{eq:trianglegamma}), as a function of $M_{\rm inv}(K^- f_0/K^- a_0)$, for (a) $m_{f_0} = 989$~MeV, (b) $m_{f_0} = 987$~MeV, (c) $m_{f_0} = 981$~MeV, (d) $m_{a_0} = 989$~MeV, (e) $m_{a_0} = 987$~MeV, and (f) $m_{a_0} = 981$~MeV.
Note that Eq.~(\ref{eq:trianglegamma}) is for $f_0$ production, and for $a_0$ production one multiplies it by $(g_{a_0,K^+K^-}/g_{f_0,K^+K^-})^2$.
}
\label{fig:trianglegamma}
\end{center}
\end{figure*}

Then we show ${1\over\Gamma_{J/\psi}}{d \Gamma_{J/\psi \rightarrow K^+ K^- f_0(a_0)} \over d M_{\rm inv}(K^- f_0/K^- a_0)}$, the differential branching ratio of the $J/\psi \rightarrow K^+ K^- f_0(980)(a_0(980))$ decays defined in Eq.~(\ref{eq:trianglegamma}), as a function of $M_{\rm inv}(K^- f_0/K^- a_0)$ in Fig.~\ref{fig:trianglegamma}.
We plot the results for three selected masses of $f_0(a_0)$, 989~MeV, 987~MeV, and 981~MeV. The results for $f_0$ or $a_0$ production differ only in a factor because of the different couplings $g_{f_0,K^+K^-}$ or $g_{a_0,K^+K^-}$. We observe a peak in $d \Gamma_{J/\psi \rightarrow K^+ K^- f_0(a_0)} \over d M_{\rm inv}(K^- f_0/K^- a_0)$ around $M_{\rm inv}(K^- f_0/K^- a_0) = 1515$~MeV. The peak is clear for the first mass of 989 MeV, but gradually the upper part of the spectrum falls down more slowly. This is due to the factor $\widetilde p^{\,\prime3}_{K^-}$ in Eq.~(\ref{eq:trianglegamma}), which raises fast as $M_{\rm inv}(K^- f_0/K^- a_0)$ increases. If we remove this factor the peaks are sharper. Next we integrate $d \Gamma_{J/\psi \rightarrow K^+ K^- f_0(a_0)} \over d M_{\rm inv}(K^- f_0/K^- a_0)$ over $M_{\rm inv}(K^- f_0/K^- a_0)$ from $M_{\rm inv}(K^- f_0/K^- a_0)_{\rm min} = m_{K^-} + m_{f_0(a_0)}$ to \\
$M_{\rm inv}(K^- f_0/K^- a_0)_{\rm max} = m_{J/\psi} - m_{K^+}$ to obtain
\begin{eqnarray}
\renewcommand{\arraystretch}{1.2}
\begin{array}{cc}
\hline \hline
~~~~~~~m_{f_0} ({\rm MeV})~~~~~~~& {\rm Br}(J/\psi \rightarrow K^+ K^- f_0(980))
\\ \hline
   989 &  1.56 \times 10^{-5}
\\ 987 &  1.66 \times 10^{-5}
\\ 981 &  1.37 \times 10^{-5}
\\ 989 &  3.55 \times 10^{-5}
\\ 987 &  3.79 \times 10^{-5}
\\ 981 &  3.12 \times 10^{-5}
\\ \hline \hline
\end{array}
\label{result1:f0981}
\end{eqnarray}
We can see that the results for the integrated branching ratios depend on the mass assumed for the $f_0(a_0)$ resonance. For the same mass, the $f_0$ or $a_0$ production rates differ by the ratio of the square of their couplings to $K^+ K^-$.

%
%

\begin{figure*}[hbtp]
\begin{center}
\includegraphics[width=0.4\textwidth]{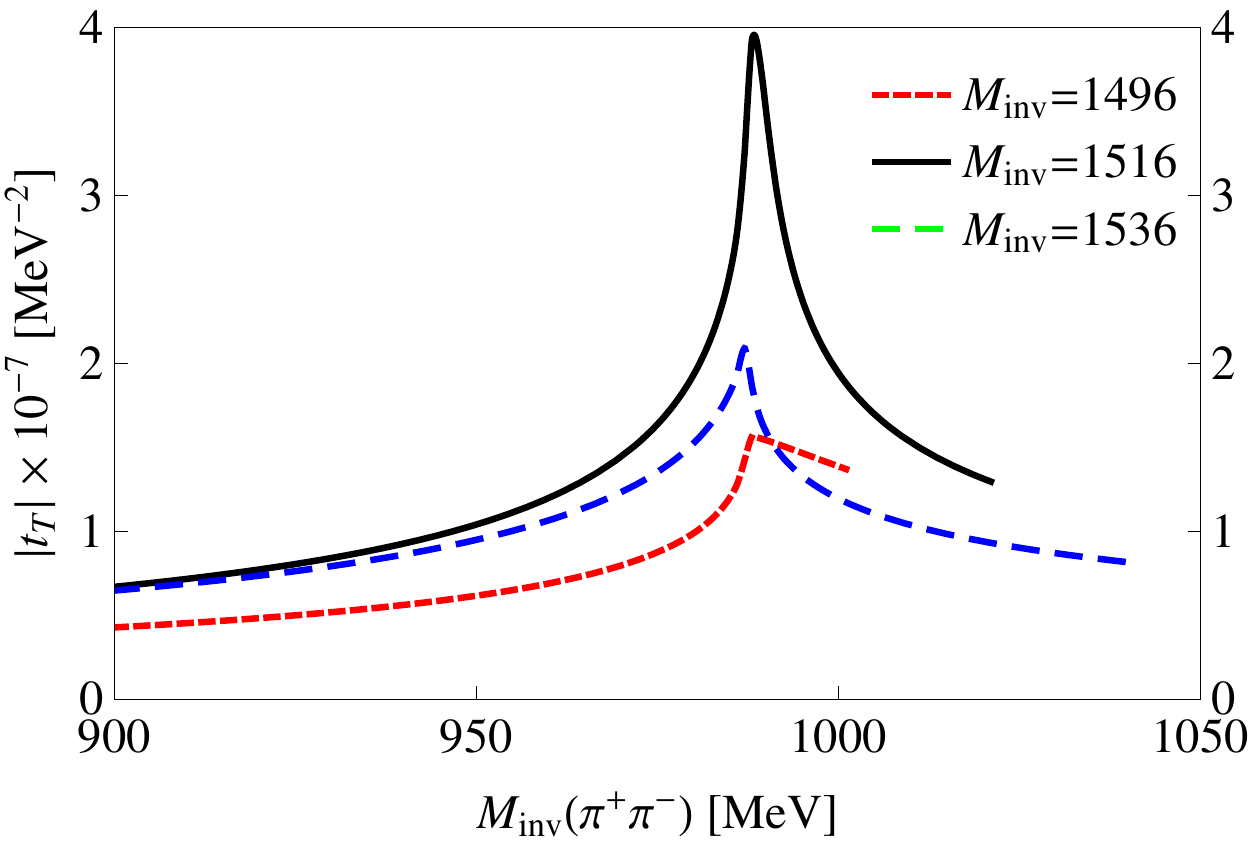}
\caption{The triangle loop $|t_T^\prime|$ for the $J/\psi \rightarrow K^+ K^- f_0(980)\rightarrow K^+ K^- \pi^+\pi^-$ reaction, defined in Eq.~(\ref{eq:fourfinal}), as a function of $M_{\rm inv}(\pi^+\pi^-)$. The red, black, and blue curves are obtained by setting $M_{\rm inv}(K^-f_0) = $ 1496 MeV, 1516 MeV, and 1536 MeV, respectively. The triangle loop $|t_T^\prime|$ for the $J/\psi \rightarrow K^+ K^- a_0(980)\rightarrow K^+ K^-\pi^0\eta$ reaction is the same as the this one.
Curves are stopped at the end of phase space for the production of $m_K$ and $M_{\rm inv}(\pi^+\pi^-)$ with the invariant mass $M_{\rm inv}(K^-f_0)$, where $t^\prime$ of Eq.~(\ref{eq:fourfinal}) vanishes.
}
\label{fig:fourdiagram}
\end{center}
\end{figure*}

In view of the changing shape and strength of the results on the mass assumed for the $f_0(a_0)$ resonance, we apply next the method discussed in Sec.~\ref{sec:fourfinal}, taking into account the mass distribution of the $f_0(980)$ and $a_0(980)$ reflected by the $t_{K^+K^-,\pi^+\pi^-}$ and $t_{K^+K^-,\pi^0\eta}$ amplitudes.
In Fig.~\ref{fig:fourdiagram} we plot $|t_T^\prime|$, the triangle loop defined in Eq.~(\ref{eq:fourfinal}) for the $J/\psi \rightarrow K^+ K^- f_0(980)\rightarrow K^+ K^- \pi^+\pi^-$ reaction, as a function of $M_{\rm inv}(\pi^+\pi^-)$ by fixing $M_{\rm inv}(K^-f_0)=$ 1496 MeV, 1516 MeV, and 1536 MeV. The distribution gets its largest strength when $M_{\rm inv}(K^-f_0)$ is near 1516 MeV. The triangle loop $|t_T^\prime|$ for the $J/\psi \rightarrow K^+ K^- a_0(980)\rightarrow K^+ K^- \pi^0\eta$ reaction is the same as this one. The $|t_T^\prime|$ function is stopped at the $M_{\rm inv}(\pi^+\pi^-)$ which makes $k^\prime$ of Eq.~(\ref{eq:k}) zero. Note that $t^\prime$ of Eq.~(\ref{eq:fourfinal}), proportional to $k^\prime$, vanishes in this point, where one has the frontier of the phase space.

%
%

\begin{figure*}[hbtp]
\begin{center}
\subfigure[]{
\includegraphics[width=0.4\textwidth]{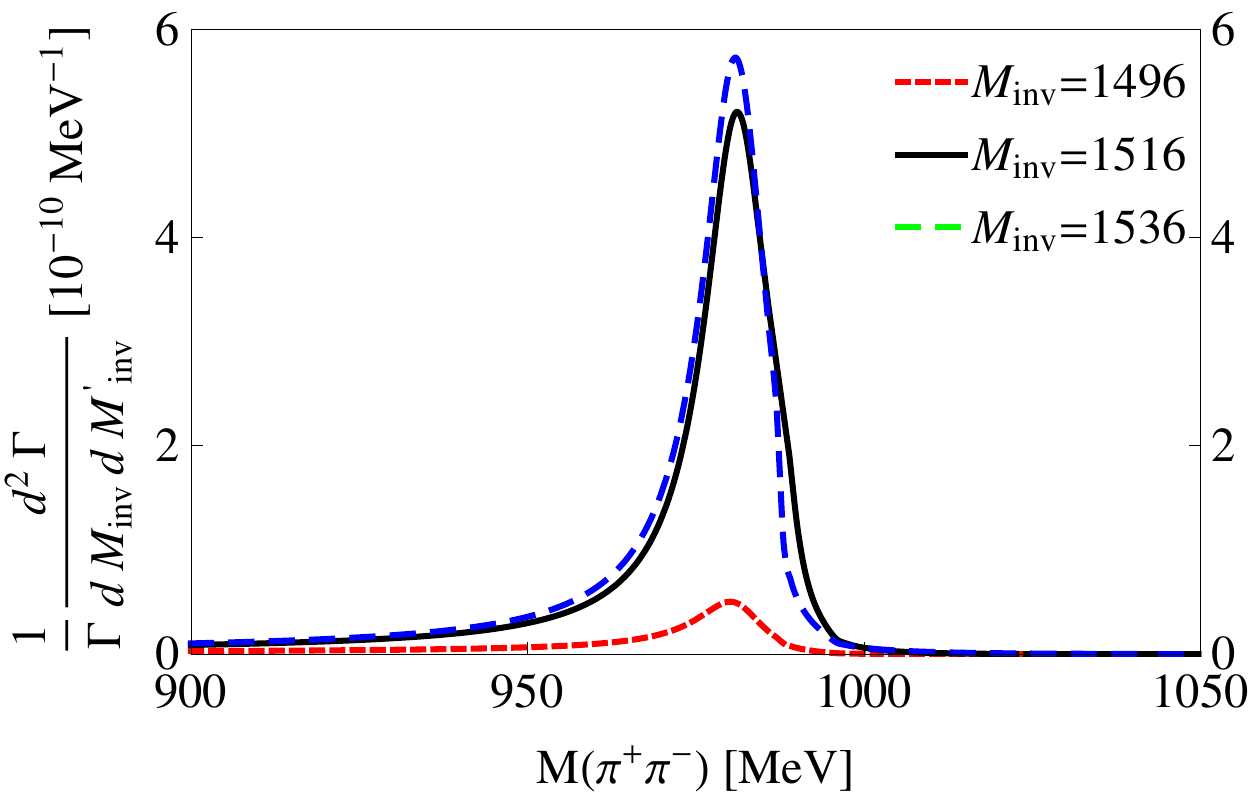}}
\subfigure[]{
\includegraphics[width=0.4\textwidth]{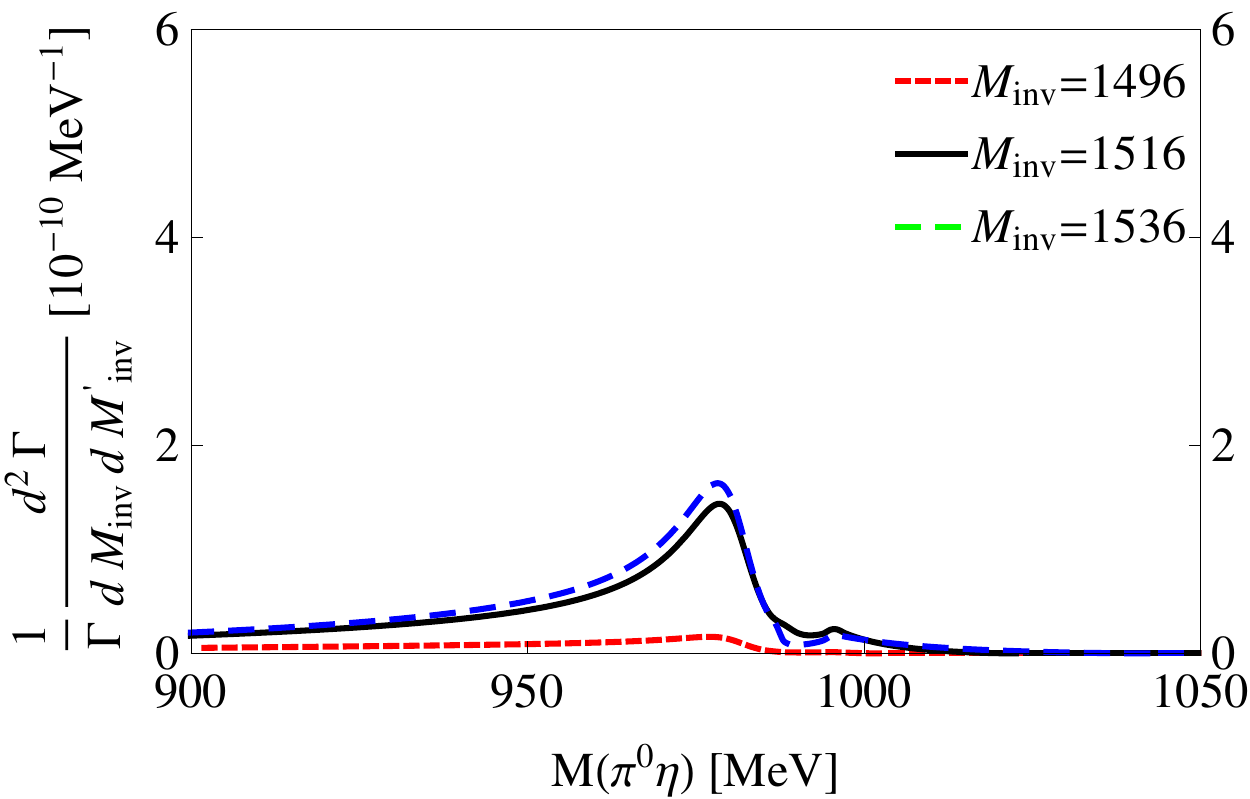}}
\caption{(a) ${1 \over \Gamma_{J/\psi}}{d^2 \Gamma_{J/\psi \rightarrow K^+K^- f_0\rightarrow K^+ K^- \pi^+\pi^-} \over d M_{\rm inv}(K^-f_0)dM_{\rm inv}(\pi^+\pi^-)}$ as a function of $M_{\rm inv}(\pi^+\pi^-)$.
(b) ${1 \over \Gamma_{J/\psi}}{d^2 \Gamma_{J/\psi \rightarrow K^+K^- a_0\rightarrow K^+ K^- \pi^0 \eta} \over d M_{\rm inv}(K^-f_0)dM_{\rm inv}(\pi^0 \eta)}$ as a function of $M_{\rm inv}(\pi^0\eta)$. The red, black, and blue curves are obtained by setting $M_{\rm inv}(K^-f_0/K^-a_0) = $ 1496 MeV, 1516 MeV, and 1536 MeV, respectively.
}
\label{fig:fourgamma}
\end{center}
\end{figure*}

\begin{figure*}[hbtp]
\begin{center}
\subfigure[]{
\includegraphics[width=0.4\textwidth]{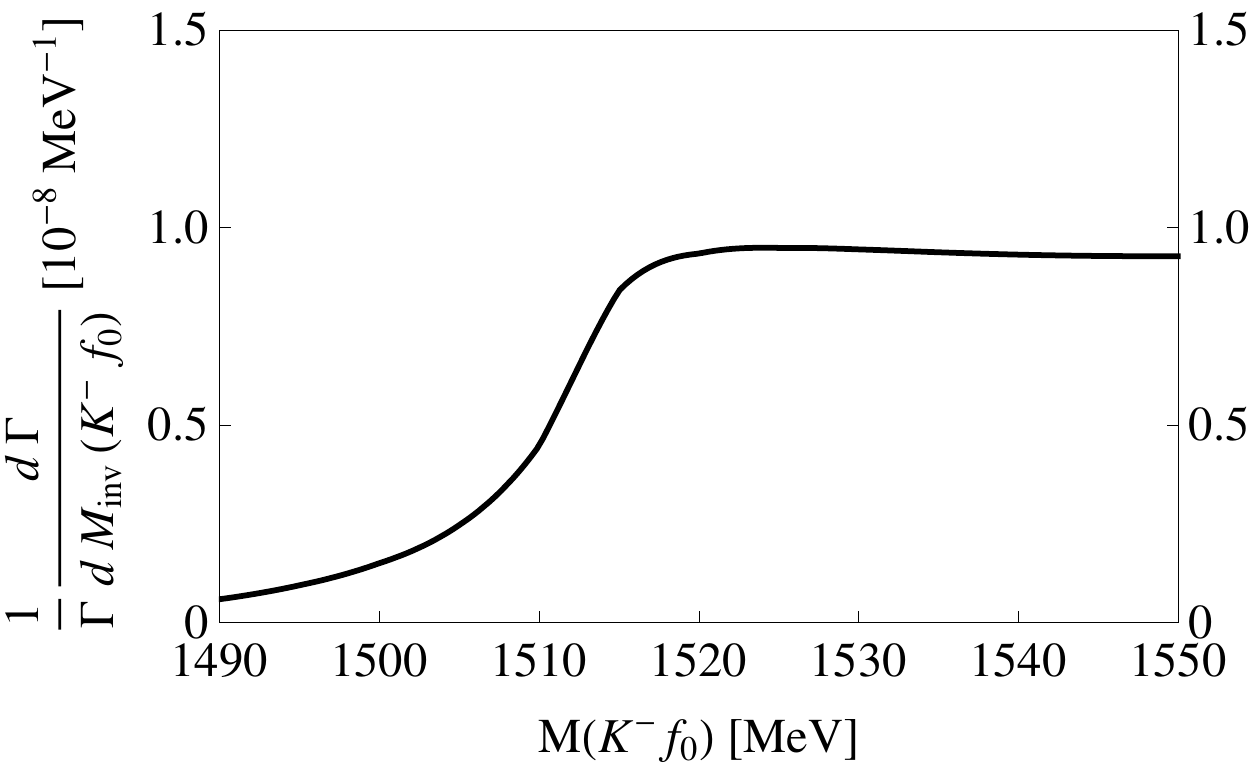}}
~~~~~~~~~~
\subfigure[]{
\includegraphics[width=0.4\textwidth]{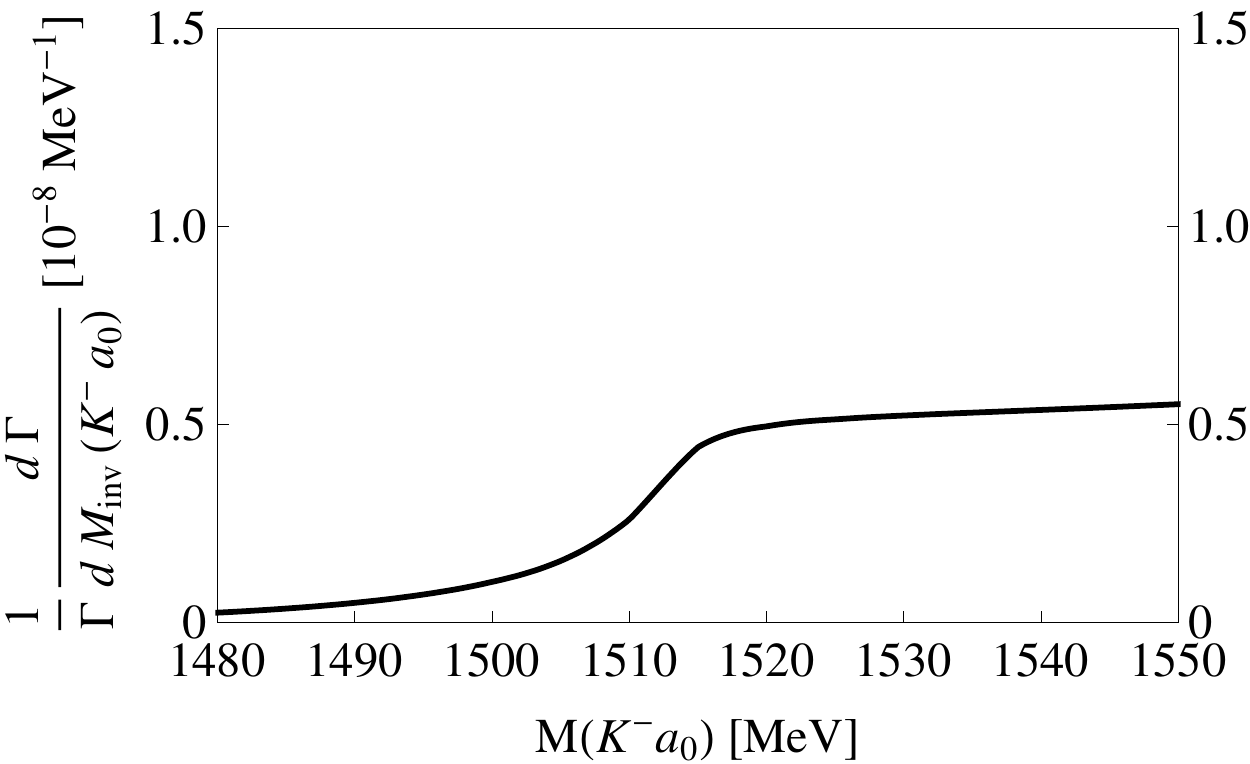}}
\caption{(a) the differential branching ratio ${1 \over \Gamma_{J/\psi}}{d \Gamma_{J/\psi \rightarrow K^+K^- f_0\rightarrow K^+ K^- \pi^+\pi^-} \over d M_{\rm inv}(K^-f_0)}$, defined in Eq.~(\ref{eq:fourgamma1}), as a function of $M_{\rm inv}(K^-f_0)$.
(b) the differential branching ratio ${1 \over \Gamma_{J/\psi}}{d \Gamma_{J/\psi \rightarrow K^+K^- a_0\rightarrow K^+ K^- \pi^0 \eta} \over d M_{\rm inv}(K^-a_0)}$, defined in Eq.~(\ref{eq:fourgamma2}), as a function of $M_{\rm inv}(K^- a_0)$.}
\label{fig:fourgamma1}
\end{center}
\end{figure*}

We also show ${1\over\Gamma_{J/\psi}}{d^2 \Gamma_{J/\psi \rightarrow K^+ K^- f_0(980)\rightarrow K^+ K^- \pi^+\pi^-} \over d M_{\rm inv}(K^- f_0)  d M_{\rm inv}(\pi^+ \pi^-)}$ in the left panel of Fig.~\ref{fig:fourgamma}, that is the double differential branching ratio of the $J/\psi \rightarrow K^+ K^- f_0(980)\rightarrow K^+ K^- \pi^+\pi^-$ reaction defined in Eq.~(\ref{eq:doublegamma}). We show this as a function of $M_{\rm inv}(\pi^+ \pi^-)$ by fixing $M_{\rm inv}(K^- f_0)=$ 1496 MeV, 1516 MeV, and 1536 MeV. A strong peak can be found when $M_{\rm inv}(\pi^+\pi^-)$ is around 980 MeV, corresponding to the $f_0(980)$. Consequently, most of the contribution comes from $M_{\rm inv}(\pi^+\pi^-) \in [900, \, 1050]$~{\rm MeV}, and we can restrict the integral in $M_{\rm inv}(\pi^+\pi^-)$ to this region when calculating the mass distribution ${1\over\Gamma_{J/\psi}}{d \Gamma_{J/\psi \rightarrow K^+ K^- f_0(980)\rightarrow K^+ K^- \pi^+\pi^-} \over d M_{\rm inv}(K^- f_0)}$.
Unlike in Fig.~\ref{fig:fourdiagram}, the strength for $M_{\rm inv}(K^- f_0) = 1536$~MeV is a bit bigger than that for 1516~MeV. This is because of the factor $\left(\widetilde p^{\,\prime\prime}_{K^-}\right)^3$ in Eq.~(\ref{eq:doublegamma}).
Similarly, we show \\
${1\over\Gamma_{J/\psi}}{d^2 \Gamma_{J/\psi \rightarrow K^+ K^- a_0(980)\rightarrow K^+ K^- \pi^0\eta} \over d M_{\rm inv}(K^- a_0)  d M_{\rm inv}(\pi^0 \eta)}$ in the right panel of Fig.~\ref{fig:fourgamma}. Again, we can restrict $M_{\rm inv}(\pi^0\eta)$ to the region $M_{\rm inv}(\pi^0\eta) \in [900, \, 1050]$~{\rm MeV}, and perform the integration
\begin{eqnarray}
\nonumber && {1\over\Gamma_{J/\psi}}{d \Gamma_{J/\psi \rightarrow K^+ K^- f_0(980)\rightarrow K^+ K^- \pi^+\pi^-} \over d M_{\rm inv}(K^- f_0)}
\\ \nonumber &=&
{1\over\Gamma_{J/\psi}} \int_{900~{\rm MeV}}^{1050~{\rm MeV}} d M_{\rm inv}(\pi^+ \pi^-)
\\ && ~~~~~~~~~~ \times {d^2 \Gamma_{J/\psi \rightarrow K^+ K^- f_0(980)\rightarrow K^+ K^- \pi^+\pi^-} \over d M_{\rm inv}(K^- f_0)  d M_{\rm inv}(\pi^+ \pi^-)} \, ,
\label{eq:fourgamma1}
\\ \nonumber && {1\over\Gamma_{J/\psi}}{d \Gamma_{J/\psi \rightarrow K^+ K^- a_0(980)\rightarrow K^+ K^- \pi^0\eta} \over d M_{\rm inv}(K^- a_0)}
\\ \nonumber &=&
{1\over\Gamma_{J/\psi}} \int_{900~{\rm MeV}}^{1050~{\rm MeV}} d M_{\rm inv}(\pi^0 \eta)
\\ && ~~~~~~~~~~ \times {{d^2 \Gamma_{J/\psi \rightarrow K^+ K^- a_0(980)\rightarrow K^+ K^- \pi^0\eta} \over d M_{\rm inv}(K^- a_0) d M_{\rm inv}(\pi^0 \eta)}} \, .
\label{eq:fourgamma2}
\end{eqnarray}
The (single) differential branching ratios obtained are shown in Fig.~\ref{fig:fourgamma1}. We see a clear structure around 1515~MeV coming from the peak of the triangle loop $t_T^\prime$, but we also observe strong contribution from the larger $K^- f_0/K^- a_0$ invariant masses produced by the $\left(\widetilde p^{\,\prime\prime}_{K^-}\right)^3$ factor of Eq.~(\ref{eq:doublegamma}). Finally, we integrate from $M_{\rm inv}(K^- f_0/K^- a_0)_{\rm min} = m_{K^-} + m_{f_0(a_0)}$ to $M_{\rm inv}(K^- f_0/K^- a_0)_{\rm max} = m_{J/\psi} - m_{K^+}$ to obtain
\begin{eqnarray}
\nonumber {\rm Br}(J/\psi \rightarrow K^+ K^- f_0(980)\rightarrow K^+ K^- \pi^+\pi^-) &=& 7.6 \times 10^{-6} \, ,
\\ \nonumber {\rm Br}(J/\psi \rightarrow K^+ K^- a_0(980) \rightarrow K^+ K^- \pi^0 \eta) &=& 5.2 \times 10^{-6} \, .
\\ \label{result2:f0}
\end{eqnarray}
These rates should be multiplied by two to account for the mechanisms of Fig~\ref{fig:fourfinal}(b) and Fig~\ref{fig:fourfinal}(d), if one looks for $J/\psi \rightarrow K^+ K^- f_0(980)(a_0(980))$ independently of which of the $K$'s is the fast one.

%
%

As we can see, the explicit consideration of the $t_{K^+K^-,\pi^+\pi^-}$ and $t_{K^+K^-,\pi^0\eta}$ changes the final shape of the differential width and integrated branching ratio. We should note that in the case of the $f_0$ production the method of Sec.~\ref{sec:triangle} accounts for all the decay modes of the $f_0$, $\pi^+\pi^-$ and $\pi^0\pi^0$, the latter with a strength $1/2$ compared to the one of $\pi^+\pi^-$. To compare the result of Eq.~(\ref{result2:f0}) with those of Eq.~(\ref{result1:f0981}) we must multiply the result of Eq.~(\ref{result2:f0}) by $3/2$ and then the results are more similar. The discrepancies are bigger in the case of the $a_0$ production. This should not be a surprise since the $a_0(980)$ is a border line state between a bound $K \bar K$ state and a threshold cusp, as a consequence of which the coupling of $a_0$ to $K^+K^-$ has large uncertainties, in which case, the method used in Sec.~\ref{sec:fourfinal} is more reliable.

%
%

\section{Conclusion}
\label{sec:conclusion}

We have studied the $J/\psi \rightarrow K^+ K^- f_0(980)(a_0(980))$ decays and have seen that they are driven by a triangle singularity, peaking at $M_{\rm inv}(K^- f_0/K^- a_0) \approx 1515$~MeV. The process proceeds as follows: In a first step the $J/\psi$ decays to $K^+ K^- \phi$. The $K^+$ and $K^-$ momenta are very distinct in the process and we select for our study the mode with $K^+$ with large momentum and $K^-$ with small momentum. The opposite case provides the same contribution. In a second step the $\phi$ decays into $K^+ K^-$ and the primary $K^-$ together with the $K^+$ from the $\phi$ decay merge to give the $f_0(980)$ or $a_0(980)$ resonance. The mechanism implicitly assumes that the $f_0(980)$ and $a_0(980)$ resonances are not produced directly but are a consequence of the final state interaction of the $K^+ K^-$. This is the basic finding of the chiral unitary approach where these two resonances are the consequence of the pseudoscalar-pseudoscalar interaction in coupled channels and not $q \bar q$ objects.

Using as input empirical information from the $J/\psi \rightarrow K^+ K^- \phi$ decay, we are able to determine the double differential decay width in terms of the $K^- f_0/K^-a_0$ invariant mass and the $\pi^+\pi^-/\pi^0\eta$ from the decay of the $f_0(980)/a_0(980)$, respectively. We find very distinct shapes of the double differential distributions, and the single one, $d\Gamma / dM_{\rm inv}(K^- f_0/K^- a_0)$, with a sharp raise of this magnitude around $M_{\rm inv}(K^- f_0/K^- a_0) \approx 1515$~MeV, where the triangle singularity appears. All these features are tied to the nature of the $f_0(980)$ and $a_0(980)$ resonances as dynamically generated from the interaction of pseudoscalar mesons, and its experimental observation should bring valuable information on the important issue of the nature of the low-lying scalar mesons. We find the branching ratios of the order of $10^{-5}$, which are accessible in present facilities, where many $J/\psi$ branching ratios of the order of $10^{-6}\sim10^{-7}$ have already been measured.

%
\section*{Acknowledgments}
%

This work is partly supported by
the National Natural Science Foundation of China under Grants Nos. 11505158, 11565007, 11722540, and 11847317,
the Fundamental Research Funds for the Central Universities,
and
the Academic Improvement Project of Zhengzhou University.
This work is also partly supported by
the Spanish Ministerio de Economia y Competitividad
and
European FEDER funds under the contract number FIS2011-28853-C02-01, FIS2011-28853-C02-02, FIS2014-57026-REDT, FIS2014-51948-C2-1-P, and FIS2014-51948-C2-2-P.

\end{document}